\documentclass[sigconf]{acmart}

\usepackage{amsmath,amsfonts}
\usepackage{algpseudocode}
\usepackage{graphicx}
\usepackage{textcomp}
\usepackage{xcolor}
\usepackage{enumitem}
\usepackage{multirow}
\usepackage{array}
\usepackage{amsthm}
\usepackage{adjustbox}
\usepackage[normalem]{ulem}

\AtBeginDocument{%
  \providecommand\BibTeX{{%
    \normalfont B\kern-0.5em{\scshape i\kern-0.25em b}\kern-0.8em\TeX}}}

\setcopyright{acmcopyright}
\copyrightyear{2022}
\acmYear{2022}
\acmDOI{XXXXXXX.XXXXXXX}

\acmConference[AFT '22]{AFT '22: Conference on Advances in Financial Technologies}{September 19--21, 2022}{Cambridge, MA}
\acmPrice{15.00}
\acmISBN{978-1-4503-XXXX-X/18/06}

\begin{document}

\title{Atomic cross-chain exchanges of shared assets}

\author{Krishnasuri Narayanam}
\affiliation{%
  \country{IBM Research, India}}
\email{knaraya3@in.ibm.com}

\author{Venkatraman Ramakrishna}
\affiliation{%
  \country{IBM Research, India}}
\email{vramakr2@in.ibm.com}

\author{Dhinakaran Vinayagamurthy}
\affiliation{%
  \country{IBM Research, India}}
\email{dvinaya1@in.ibm.com}

\author{Sandeep Nishad}
\affiliation{%
  \country{IBM Research, India}}
\email{sandeep.nishad1@ibm.com}


\newtheorem{claim}{Claim}

\begin{abstract}
  A core enabler for blockchain or DLT interoperability is the ability to atomically exchange assets held by mutually untrusting owners on different ledgers. This atomic swap problem has been well-studied, with the Hash Time Locked Contract ($\mathsf{HTLC}$) emerging as a canonical solution. $\mathsf{HTLC}$ ensures atomicity of exchange, albeit with caveats for node failure and timeliness of claims. But a bigger limitation of $\mathsf{HTLC}$ is that it only applies to a model consisting of two adversarial parties having sole ownership of a single asset in each ledger. Realistic extensions of the model in which assets may be jointly owned by multiple parties, all of whose consents are required for exchanges, or where multiple assets must be exchanged for one, are susceptible to collusion attacks and hence cannot be handled by $\mathsf{HTLC}$. In this paper, we generalize the model of asset exchanges across DLT networks and present a taxonomy of use cases, describe the threat model, and propose $\mathsf{MPHTLC}$, an augmented $\mathsf{HTLC}$ protocol for atomic multi-owner-and-asset exchanges. We analyze the correctness, safety, and application scope of $\mathsf{MPHTLC}$. As proof-of-concept, we show how $\mathsf{MPHTLC}$ primitives can be implemented in networks built on Hyperledger Fabric and Corda, and how $\mathsf{MPHTLC}$ can be implemented in the Hyperledger Labs Weaver framework by augmenting its existing $\mathsf{HTLC}$ protocol.
\end{abstract}




\maketitle

\section{Introduction}
\label{sec:introduction}
The existence of diverse blockchain and distributed ledger technologies (DLTs) and networks, especially permissioned ones, have spurred research \cite{ABGHKNPRVMIDDLEWARE19, BVGCACM22}, development \cite{Cactus20, Weaver21}, and standardization \cite{tc307, IETFGatewaysDraft} efforts in interoperability. If networks cannot interoperate, their business processes (contracts) cannot interlink and their assets get trapped in silos, preventing them from scaling up and decreasing their relevance in the overall blockchain economy~\cite{IETFGatewaysDraft, Weaver21}. A key interoperation enabler is the \textit{atomic swap}~\cite{AtomicSwap}, or the ability to exchange assets in two different ledgers/networks (\textit{we will use these terms interchangeably for systems that manage shared assets using blockchain or DLT}) between a pair of owners atomically; i.e., the exchange happens or both ledgers revert to their original states. This problem has gained increased salience with the emergence of Decentralized Finance (DeFi)~\cite{DeFi} and Central Bank Digital Currencies (CBDCs)~\cite{cbdc}. But, as we will see, the state-of-the-art, and existing models and solutions, are inadequate for complex exchanges involving multiple owners and assets, which require new solutions.

The canonical asset swap can be understood through an example. Alice and Bob possess accounts in both the Bitcoin and Ethereum Main networks. They come to an (off-chain) agreement whereby Alice will give Bob 10 BTC in exchange for, say, 12 ETH. The expected outcome of this exchange/swap is that a transfer of 10 BTC from Alice to Bob is confirmed on the Bitcoin network while simultaneously a transfer of 12 ETH from Bob to Alice is confirmed on the Ethereum Mainnet. Both transfers occur or neither does, ideally without requiring a trusted mediator or complex cross-network synchronization (e.g., linking clocks, coordinating block confirmation frequencies). A similar example can be conceived between two permissioned networks, built on, e.g. Hyperledger Fabric and Corda, using tokens designed for those DLT platforms~\cite{FabricTokens, CordaTokens}. \textit{Delivery-vs-payment} (DvP), a common DeFi use case, offers additional motivation. Consider a DLT network where different commercial banks possess retail accounts of a Central Bank Digital Currency (CBDC) and another DLT network where banks allow investors to trade securities (e.g. bonds). The transfer of a bond on the latter must simultaneously accompany a CBDC payment on the former ~\cite{cbdcbondinterop}.

\begin{figure*}[ht]
    \centering
    \includegraphics[width=0.8\textwidth]{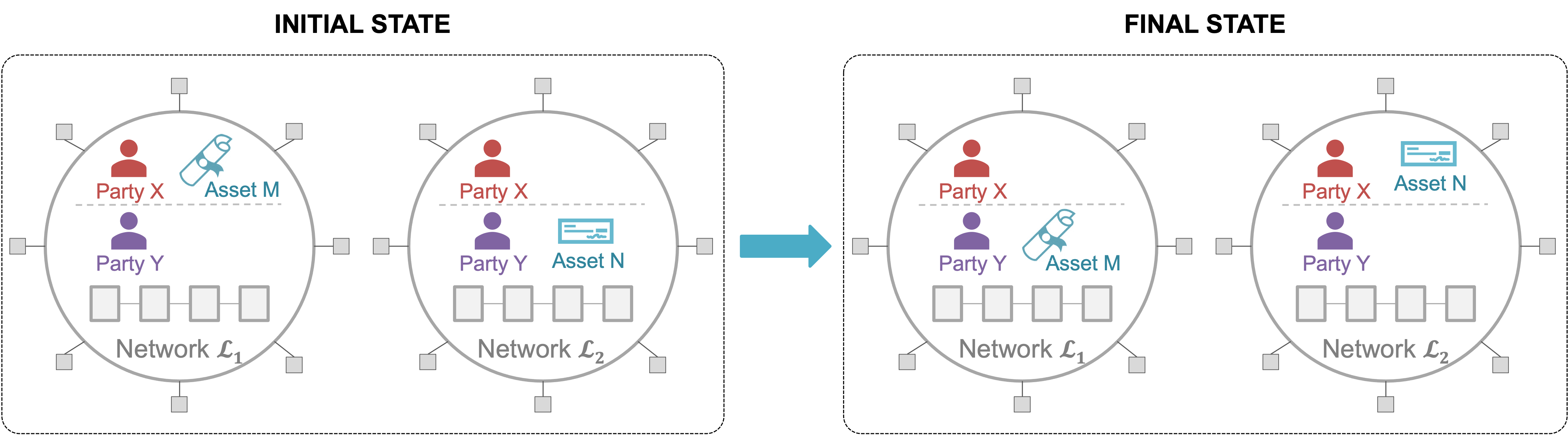}
    \caption{Asset Exchange (Atomic Swap) Model}
    \label{fig:asset-exchange-model}
\end{figure*}

\begin{figure*}[ht]
    \centering
    \includegraphics[width=0.65\textwidth]{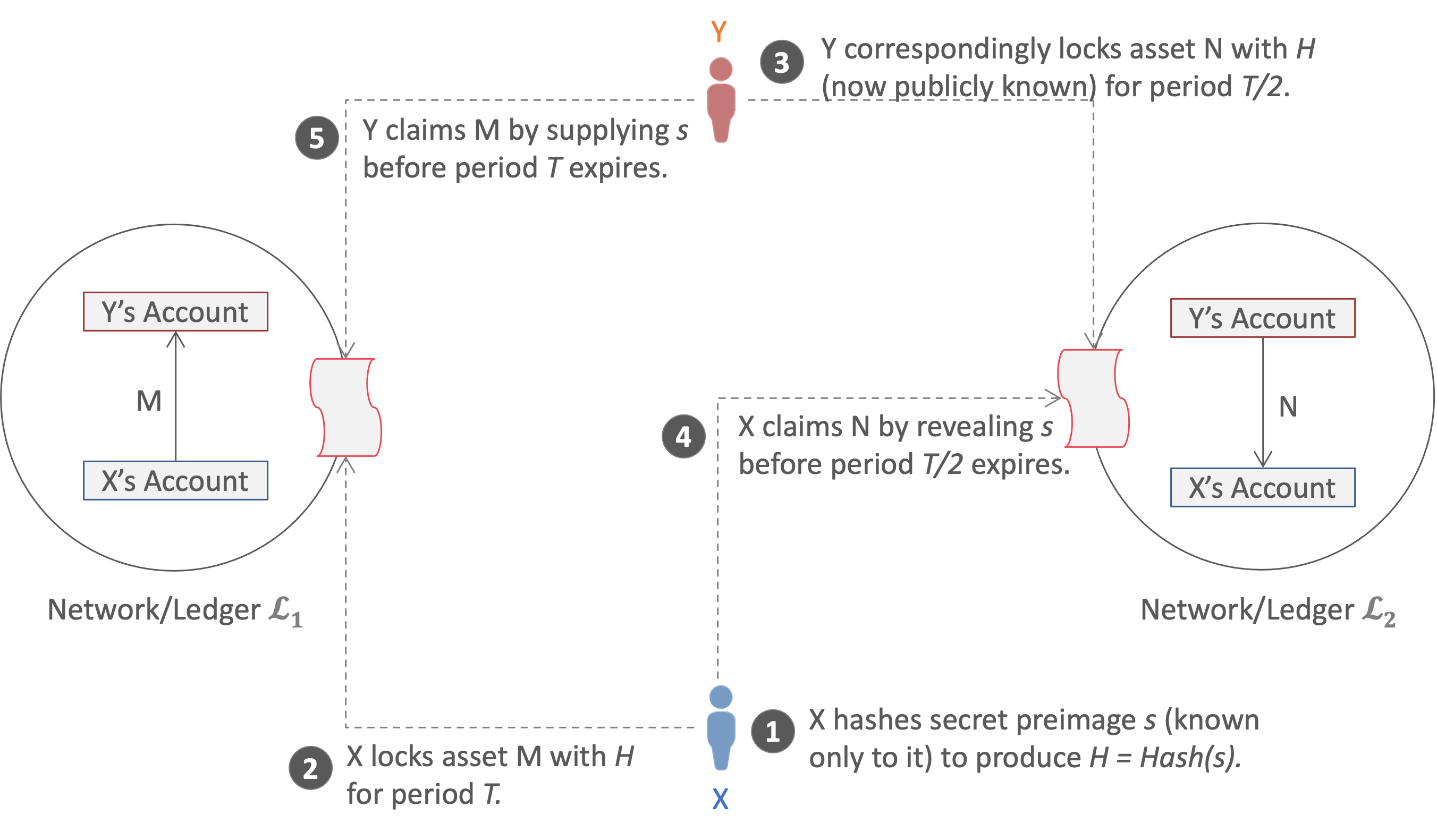}
    \caption{Hash Time Locked Contract ($\mathsf{HTLC}$) Protocol}
    \label{fig:htlc}
\end{figure*}

From these examples, we can extrapolate a general model of exchanging an asset $M$, owned by party $X$ in ledger $\mathcal{L}_1$, for asset $N$, owned by party $Y$ in  ledger $\mathcal{L}_2$ (Figure~\ref{fig:asset-exchange-model}). The canonical solution to complete such atomic swaps is the \textit{Hash Time Locked Contract} ($\mathsf{HTLC}$) \cite{NolanSwap13}. This describes both a contract that supports the locking and claiming of an asset within a fixed time duration and a protocol (see Figure~\ref{fig:htlc}), whose "happy path" is as follows:
\begin{itemize}[leftmargin=*]
    \item $X$ hashes secret $s$ to produce $H = Hash(s)$
    \item $X$ locks $M$ on $\mathcal{L}_1$, specifying $Y$ as the recipient, with $H$ as guard for duration $T$ using a blockchain or smart contract transaction 
    \item $Y$ similarly locks $N$ on $\mathcal{L}_2$ with $H$ (now publicly known), specifying $X$ as the recipient, for duration $T/2$
    \item $X$ claims $N$ on $\mathcal{L}_2$ by revealing $s$ (which, when hashed, must match $H$) via a transaction within $T/2$
    \item $Y$ similarly claims $M$ on $\mathcal{L}_1$ by supplying $s$ (now publicly known) after $X$'s claim but within $T$
\end{itemize}
This protocol works because (i) both $X$ and $Y$ have visibility into both ledgers, be they in public or permissioned networks, (ii) $X$'s secret is revealed to $Y$ only after both assets are locked, (iii) the secret is revealed in public within a transaction for any claimers to use, (iv) $Y$ has sufficient time to make a claim after $X$ does so by revealing $s$; if $Y$ locks immediately after $X$ does, $X$ has up to $T/2$ to make its claim, leaving $Y$ at least $T/2$ before the expiration of $X$'s lock, and (v) only $X$ and $Y$ will be able to claim assets as designated recipients. To support $\mathsf{HTLC}$, a DLT platform must enable asset locking (or enforce any general constraint) for a fixed time duration.

\begin{figure*}[ht]
    \centering
    \includegraphics[width=0.75\textwidth]{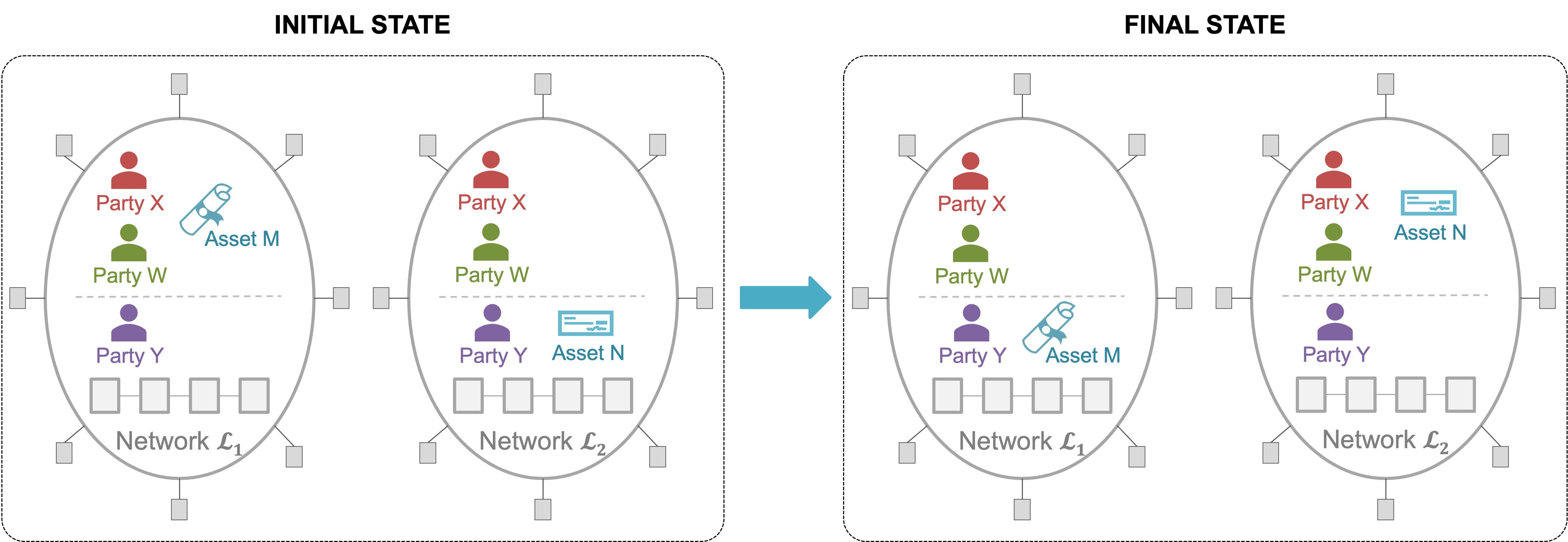}
    \caption{Jointly Owned Asset Exchange}
    \label{fig:asset-exchange-variation-1}
\end{figure*}

\begin{figure*}[ht]
    \centering
    \includegraphics[width=0.75\textwidth]{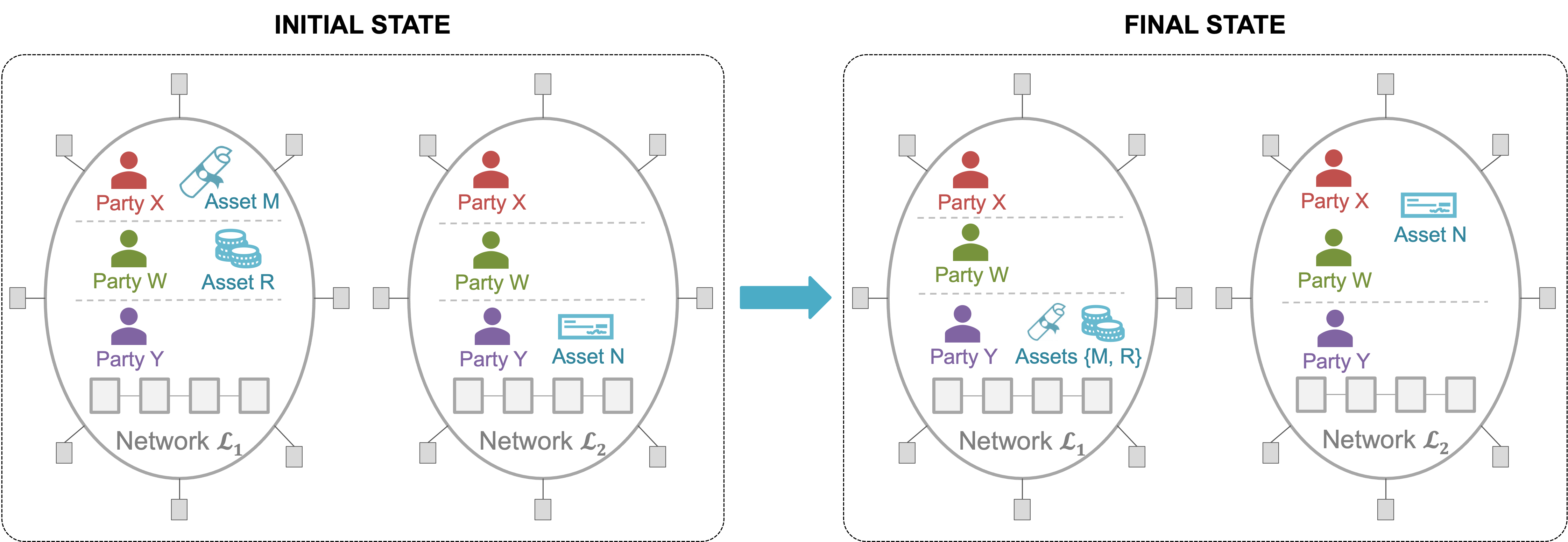}
    \caption{Multi-Asset Exchange}
    \label{fig:asset-exchange-variation-2}
\end{figure*}

Can $\mathsf{HTLC}$ be generalized for atomic exchanges of more than two assets and joint ownership (or \textit{co-ownership}) of assets by multiple parties? In one extension to the basic model, we can add party $W$ to the mix so that $X$ and $W$ co-own $M$ in $\mathcal{L}_1$ (Figure~\ref{fig:asset-exchange-variation-1}), with the desired outcome of $X$ and $W$ co-owning $N$ in $\mathcal{L}_2$ and $Y$ owning $M$ in $\mathcal{L}_1$. E.g., $M$ is a title to property owned jointly by two parties, who are willing to sell it to a third party in exchange for money being transferred to a joint account held by them. In another extension, $W$ owns a third asset $R$ in $\mathcal{L}_1$, and the desired outcome is that $Y$ acquires $M$ and $R$ in $\mathcal{L}_1$ in exchange for transferring $N$ in $\mathcal{L}_2$ jointly to $X$ and $W$ (Figure~\ref{fig:asset-exchange-variation-2}). E.g., $M$ and $R$ are separate titles to two co-located properties that are sold together to a third party in exchange for money being transferred to the owners' joint account. Though these scenarios seem superficially similar to the base case, application of the basic $\mathsf{HTLC}$ protocol in either is susceptible to attack whereby $X$ can cheat $W$ by colluding with $Y$ so that $W$ loses ownership of its asset while both $X$ and $Y$ gain something, as we will see in Section~\ref{sec:modeling}. Addressing such flaws and creating a generalized $\mathsf{HTLC}$ solution for generalized asset exchanges is the goal of this paper, whose contributions can be summarized as follows:
\begin{itemize}[leftmargin=*]
    \item A generalized model of asset exchanges across DLT networks with a taxonomy of use cases exhaustively covering all atomic cross-network transaction possibilities. We present an associated threat model, call out desired safety properties, and identify the shortcomings of $\mathsf{HTLC}$ with respect to these properties.
    \item An augmented $\mathsf{HTLC}$ protocol $\mathsf{MPHTLC}$, using multi-party computation (MPC), for atomic multi-party-and-asset exchanges that is proven secure in this model.
    \item An implementation of this protocol in the open-source Hyperledger Labs Weaver framework to facilitate generic atomic multi-party-and-asset exchanges between distributed applications built on Hyperledger Fabric and Corda networks. We also perform a simple evaluation to show that $\mathsf{MPHTLC}$ is practically usable.
\end{itemize}


In Section~\ref{sec:modeling}, we generalize the asset exchange (or atomic swap) problem, and model the scenarios and resulting threats exhaustively. Our solution for the general model that handles these threats is described in Section~\ref{sec:solution} and its efficacy is analyzed in Section~\ref{sec:analysis}. Practical design considerations, and suggestive implementations on two prominent DLT platforms, are discussed in Section~\ref{sec:implementation}. After covering related work in Section~\ref{sec:relatedWork}, we conclude with suggestions for future work in Section~\ref{sec:conclusion}.
\section{Generalizing Asset Exchanges}
\label{sec:modeling}


In this section, we will comprehensively analyze and categorize the modes of asset exchanges that a group of parties can carry out across two distributed ledgers. We will scope the problem space, define basic and complex transaction types, and present a formal generalized asset exchange ($\mathcal{GAE}$) model. 
Then we will show that conventional $\mathsf{HTLC}$ when applied to this model is prone to attacks, and identify remedies which lead to our solution.

\subsection{Asset Co-Ownership}
\label{sec:co-ownerships}
Before describing the model, we will define what we mean by co-ownership of an asset and state the properties expected from a DLT to manage such co-ownership. 
Conventionally, ownership of an asset in a blockchain or smart contract system implies sole authority to change the state of that asset, lock (freeze) its state until certain conditions are met, or transfer it to another owner. The state of an asset is subject to change according to rules defined in scripts (e.g., Bitcoin \cite{BitcoinScript}) or contracts (e.g., Ethereum \cite{EthereumContract}, Hyperledger Fabric ~\cite{FabricEuroSys18}). 
DLT systems are expected to ensure, through consensus mechanisms and well-vetted scripts/contracts, that only rightful owners can lock assets, change their properties, or transfer them.

We define co-ownership as the ability of a ledger or smart contract system to enforce similar integrity rules on assets where the owner is a collective rather than a single entity. Any state update or transfer or lock must require the consent of some or all members of the collective (i.e., co-owners). A typical way to enforce this is \textit{multisig}, or multiple signatures, ~\cite{MultiSign1,MultiSign2}, where the contract requires a quorum of signatures on a transaction to update ledger state. Our minimal expectation from a smart contract system is its ability to enforce multiple or unanimous consent on asset modifications, i.e., allow an asset to be governed by multiple users. 

\subsection{Basic Transaction Types}
\label{sec:basic-tx-types}
\begin{figure*}[ht]
    \centering
    \includegraphics[width=0.7\textwidth]{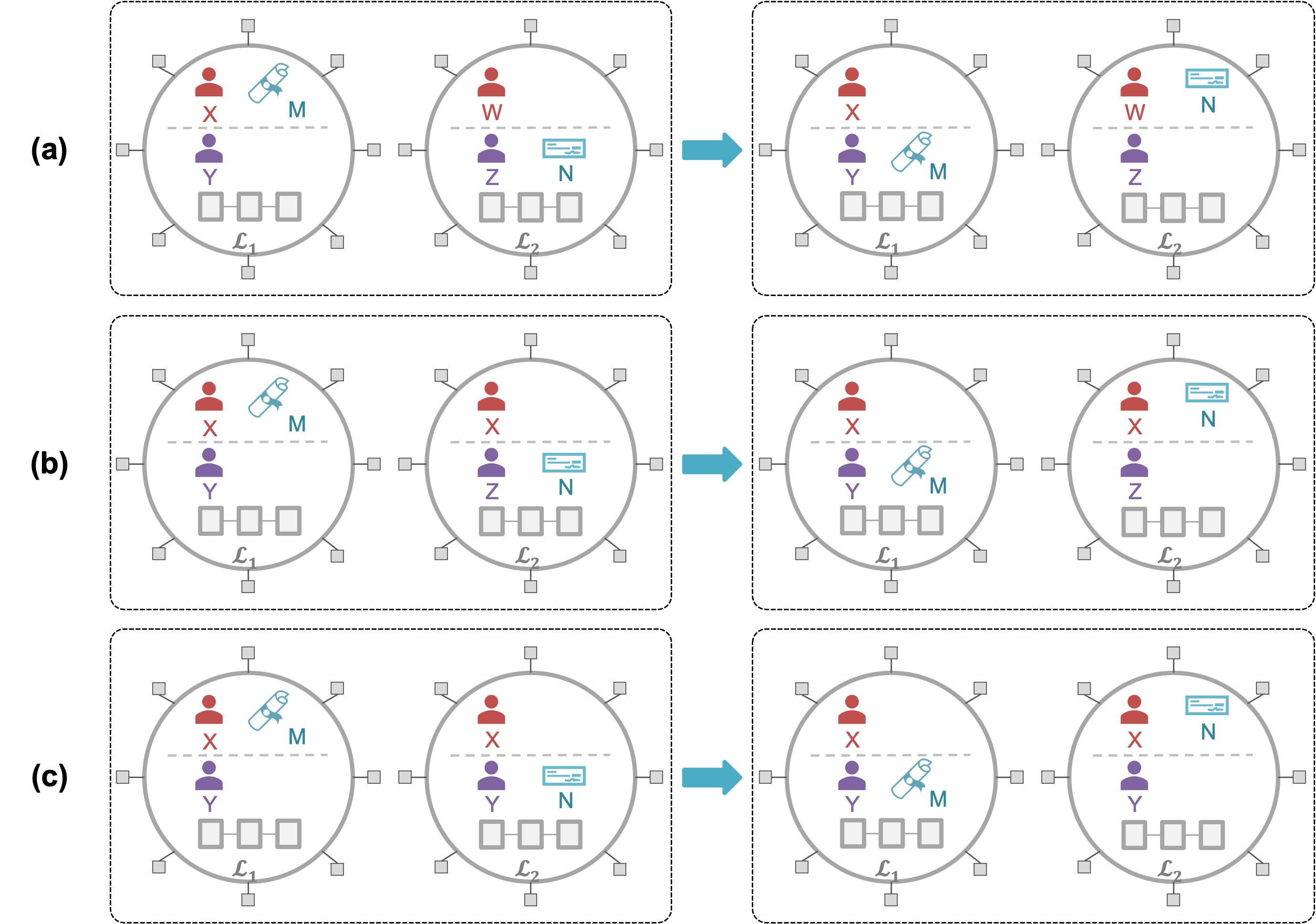}
    \caption{Basic Transaction Types: (a) Unconnected Local Transactions, (b) Cross-Ledger Replacement, (c) Cross-Ledger Swap}
    \label{fig:cross-network-tx-types}
\end{figure*}

Let us analyze \textit{cross-ledger exchange transactions} (or simply \textit{cross-ledger exchanges}), i.e., transactions that involve multiple asset ownership changes in two ledgers possibly with atomicity requirements. Figures~\ref{fig:asset-exchange-model}, ~\ref{fig:asset-exchange-variation-1}, and~\ref{fig:asset-exchange-variation-2}, illustrate examples.
We can identify and categorize patterns within cross-ledger exchanges; elements of these categories vary only in the number of co-owners and in the numbers of assets involved in the transactions. We will prove that these patterns exhaustively cover all permutations where arbitrary numbers of assets with arbitrary co-owner sets are to be exchanged across two ledgers.
To start with, we can identify certain fundamental operations on which cross-ledger exchanges are built:
\begin{itemize}[leftmargin=*]
    \item Creation of an asset with assignment of one or more co-owners
    \item Destruction of an asset with every co-owner losing ownership
    \item Changes to the co-ownership set of an asset: add co-owner(s), remove co-owner(s), or both simultaneously
\end{itemize}
Since exchange scenarios take the existence of the assets in question for granted, we can disregard the first two operations and focus on co-ownership changes. As described earlier, addition or removal of co-owners requires existing co-owners' consents, typically using signatures, though the precise consent criteria can vary with ledger or contract without changing the semantics of ownership change. 
(For example, if an asset has more than one co-owner, the replacement of one of the co-owners may or may not require consent of the other co-owners.)
Therefore, without loss of generality, we will assume in this paper that consent from co-owners is required for any addition or remove of an asset's co-owners in the context of an atomic exchange. Now we can list the cross-ledger exchanges that are built on these fundamental operations:
\begin{itemize}[leftmargin=*]
    \item \textit{Unconnected Local Transfers} (ULTs): A pair of asset transfers in two ledgers where neither the initial nor the final owner of one asset matches the initial or final owner of the other asset (as illustrated in Figure~\ref{fig:cross-network-tx-types}a). In effect, these transfers are not interdependent and can be executed simply as smart contract invocations localized to the respective ledgers. None of the involved parties have an atomicity requirement.
    \item \textit{Cross-Ledger Replacements} (CLRs): A pair of transfers where the initial owner of one asset is the same as the final owner of the other (as illustrated in Figure~\ref{fig:cross-network-tx-types}b). 
    \item \textit{Cross-Ledger Swaps} (CLSs): A pair of transfers where the initial owner of one asset is the same as the final owner of the other and vice versa (as illustrated in Figure~\ref{fig:cross-network-tx-types}c and also earlier in Figure~\ref{fig:asset-exchange-model}). 
\end{itemize}
A CLR can be viewed as a ULT pair with more constraints, where $X == W$, i.e., $X$ and $W$ are combined into a single logical entity. From $X$'s perspective (though not from $Y$'s or $Z$'s), the two transactions must happen atomically. Similarly, a CLS can be viewed as a CLR with more constraints, where $Y == Z$, and both $X$ and $Y$ desire atomicity. As we can see, more constraints create more atomicity requirements. Even though ULTs have no atomicity requirements, it is necessary to list them here for completion. A general cross-ledger exchange (defined later in this section) will be built on a combination and extrapolation of all these basic types, with some subset of parties desiring atomicity. 

Our classification covers the possibility of ULT participants $\{X,Y,Z,W\}$ desiring atomicity for their pair of transfers. Using $\mathsf{HTLC}$, the first locker (say $X$) can share the secret through an off-chain communication channel with the first claimer (say $W$). To the ledger/contract, this effectively combines X and W into a single logical entity, and the resulting scenario becomes congruent to a CLR. Further, if the second locker ($Z$) and second claimer ($Y$) are the same logical (or real-world) entity, the scenario morphs into a CLS. Hence, enforcing atomicity on a pair of ULTs (the most general transfer pair in a 2-ledger system) must correspond to a CLR or a CLS pattern, proving that our categorization of basic transaction types is exhaustive. 
Without an atomicity requirement, a ULT pair can simply be enforced through a pair of smart contract invocations in the respective ledgers in any sequence, while a CLS can be enforced using $\mathsf{HTLC}$. A CLR can also be enforced using $\mathsf{HTLC}$ if the shared entity ($X$ in Figure~\ref{fig:cross-network-tx-types}b) locks $M$ first and then claims $N$ from $Z$ by revealing its secret. Also, CLRs can be augmented by adding multiple co-owners in place of $X$ or by adding multiple asset owners in $\mathcal{L}_1$, that would jointly claim $N$ in $\mathcal{L}_2$ (as with the CLS augmentations in Figures~\ref{fig:asset-exchange-variation-1} and~\ref{fig:asset-exchange-variation-2}). Hence, we can treat CLRs and CLSs as similar exchange problems requiring similar solutions.

\subsection{Complex Cross-Ledger Transactions}
\label{sec:complex-cross-net-tx}

We can build arbitrarily complex transactions using the basic transactions as building blocks simply by:
\begin{itemize}[leftmargin=*]
    \item Adding more co-owners to each asset (Figure~~\ref{fig:asset-exchange-variation-1})
    \item Involving more assets in the exchange (Figure~\ref{fig:asset-exchange-variation-2})
\end{itemize}
In general, a complex transaction involves a set of arbitrary co-ownership transfers of an arbitrary set of assets within two ledgers, some of which require cross-network atomicity guarantees. We claim that this can be expressed as a union of ULTs, CLRs, and CLSs where each asset has one or more co-owners (let us call the generalized multi-co-owner variants as gULT, gCLR and gCLS respectively).
We assume that the contracts reflect the intents of the group of parties involved in the exchanges: i.e., some transactions require atomicity and others do not. We can assume, without loss of generality, that all transactions in an exchange scenario require atomicity and that the users have already excluded those that do not. Also, for any asset that is being replaced or swapped, we assume that all co-owners of those assets are actively involved in the cross-network transaction, even if some of those co-owners' roles are purely limited to one ledger. A ledger/contract can always exclude such co-owners from cross-network transactions, by allowing them to just endorse (sign) a local transfer, which the ledger can process concurrently with a complex atomic cross-ledger transaction. Our model and solution will manage just the complex cross-ledger transaction, which existing ledgers are not equipped for.

For example, let us augment the CLR model in Figure~\ref{fig:cross-network-tx-types}b to add an initial co-owner $W$ for asset M in ledger $\mathcal{L}_1$, so that $X$ and $W$ together transfer $M$ to $Y$. But in ledger $\mathcal{L}_2$, $W$ still does not take part in any transaction. We can represent this as follows:
\begin{itemize}[leftmargin=*]
    \item $\mathcal{L}_1: M: \{X,W\} \rightarrow \{Y\}$
    \item $\mathcal{L}_2: N: \{Z\} \rightarrow \{X\}$
\end{itemize}
From $W$'s perspective, the first transaction is a ULT. Hence this transaction pair is a combination of a ULT and a CLR. $W$ just needs to endorse $X$'s locking of $M$ in $Y$'s favor by signing the transaction. But the final outcome requires atomic commitments; hence we can merge this ULT into the CLR and require $W$ and $X$ to lock $M$ jointly. We encounter the exact same possibilities and outcome if we augment a CLS (Figure~\ref{fig:cross-network-tx-types}c) as follows:
\begin{itemize}[leftmargin=*]
    \item $\mathcal{L}_1: M: \{X,W\} \rightarrow \{Y\}$
    \item $\mathcal{L}_2: N: \{Y\} \rightarrow \{X\}$
\end{itemize}
$W$ plays an ancillary role by simply doing a ULT to $Y$, but that ULT can be merged with the CLS, making $W$ an active participant in the swap. Merging ULTs into atomic transactions preserves the final outcome (i.e., co-ownership changes remains the same), while making analysis and solution-building easier. Therefore, we can assume merging as default without our model losing any power.

Here is a scenario that shows how a complex transaction is a composition of basic transactions. This is illustrated in Figure~\ref{fig:complex-asset-exchange-example} (symbols will be explained later.) The following are the asset transfers (or asset ownership changes) occurring in the two ledgers:
\begin{itemize}[leftmargin=*]
    \item $\mathcal{L}_1: Currency: \{X,Y\} \rightarrow \{W,Y,Z\}$
    \item $\mathcal{L}_1: Security: \{T,U\} \rightarrow \{V\}$
    \item $\mathcal{L}_1: Diamond: \{Z\} \rightarrow \{V\}$
    \item $\mathcal{L}_2: Car: \{T\} \rightarrow \{T,U,W\}$
    \item $\mathcal{L}_2: House: \{Z\} \rightarrow \{T,X,Y\}$
\end{itemize}
By inspection, we can deconstruct this set of transactions:
\begin{itemize}[leftmargin=*]
    \item $\mathcal{L}_1: Currency: \{X,Y\} \rightarrow \{W,Y,Z\}$ and $\mathcal{L}_2: House: \{Z\} \rightarrow \{T,X,Y\}$ together comprise a gCLS from the perspective of $\{X,Y\}$ as one group of co-owners and $\{Z\}$ as another. $W$ and $T$ are simply claimants and don't play an active role in the swap, but their endorsements (i.e., digital signatures on the transfer transactions) are needed to finalize the transactions; hence, their roles can be merged into the CLS as a whole. (\textit{Note}: excluding $W$ and $T$ still leaves multiple co-owners ($\{X,Y,Z\}$) for the two assets being swapped, so this scenario is a complex swap according to our model. As we will see in Section~\ref{sec:htlc-limitations}, such an exchange cannot be handled as a simple swap using classic $\mathsf{HTLC}$.)
    \item $\mathcal{L}_1: Security: \{T,U\} \rightarrow \{V\}$ and $\mathcal{L}_2: Car: \{T\} \rightarrow \{T,U,W\}$ together comprise a gCLR from $U$'s perspective. $T$ simply gives or shares its co-ownership and $V$ and $W$ simply take co-ownerships, but their roles can be merged into the gCLR as a whole. It is up the the users whether or not to complete this gCLR independent of the above gCLS, but if they so intend, it can be enforced atomically with the gCLS, in effect merging the two exchanges into a single atomic exchange. This is the default assumption our solution will make when presented with a set of transactions (see Section~\ref{sec:solution}).
    \item $\mathcal{L}_1: Diamond: \{Z\} \rightarrow \{V\}$ is a ULT as it is unconnected to the other transactions in the set. But, if the users so intend, this can be completed atomically with the other transactions, and that is the default assumption our solution will make when presented with a set of transactions (see Section~\ref{sec:solution}).
\end{itemize}

\subsection{Generalized Asset Exchange Model}
\label{sec:generalized-asset-exchange-model}



We now formally model the problem of atomically changing co-ownership configurations of assets in two independent ledgers, where the co-owners across all assets belong to a common limited size group of parties.

\begin{definition}[generalized asset exchange: $\mathcal{GAE}$] This is a tuple $\langle \mathcal{L}_1,\mathcal{L}_2,\mathcal{P},\mathcal{A}_1,\mathcal{A}_2,\mathsf{IO}_1,\mathsf{IO}_2,\mathcal{AE}_1,\mathcal{AE}_2,\mathsf{FO}_1,\mathsf{FO}_2 \rangle$ where:
\begin{itemize}[leftmargin=*]
    \item $\mathcal{L}_1$ and $\mathcal{L}_2$ are two distinct ledgers
    \item $\mathcal{P}$ is a common set of parties with accounts in both $\mathcal{L}_1$ and $\mathcal{L}_2$
    \item $\mathcal{A}_1$ is a set of assets on $\mathcal{L}_1$
    \item $\mathcal{A}_2$ is a set of assets on $\mathcal{L}_2$
    \item $\mathsf{IO}_1$ is the initial asset ownership function in $\mathcal{L}_1$, defined as $\mathsf{IO}_1:\mathcal{A}_1 \to 2^\mathcal{P} - \{\varnothing\}$
    \item $\mathsf{IO}_2$ is the initial asset ownership function in $\mathcal{L}_2$, defined as $\mathsf{IO}_2:\mathcal{A}_2 \to 2^\mathcal{P} - \{\varnothing\}$
    \item $\mathcal{AE}_1$ is a set of assets in $\mathcal{L}_1$ to be exchanged, where $\mathcal{AE}_1 \subseteq \mathcal{A}_1$
    \item $\mathcal{AE}_2$ is a set of assets in $\mathcal{L}_2$ to be exchanged, where $\mathcal{AE}_2 \subseteq \mathcal{A}_2$
    \item $\mathsf{FO}_1$ is the final asset ownership function in $\mathcal{L}_1$ after the exchange, defined as $\mathsf{FO}_1:\mathcal{A}_1 \to 2^\mathcal{P} - \{\varnothing\}$, where:
    \begin{itemize}
        \item $\forall a \in \mathcal{A}_1 - \mathcal{AE}_1, \mathsf{FO}_1(a) = \mathsf{IO}_1(a)$, and
        \item $\forall a \in \mathcal{AE}_1, \mathsf{FO}_1(a) \in 2^{\mathcal{P}} - \{\varnothing\} \land \mathsf{FO}_1(a) \neq \mathsf{IO}_1(a)$
    \end{itemize}
    \item $\mathsf{FO}_2$ is the final asset ownership function in $\mathcal{L}_2$ after the exchange, defined as $\mathsf{FO}_2:\mathcal{A}_2 \to 2^\mathcal{P} - \{\varnothing\}$, where:
    \begin{itemize}
        \item $\forall a \in \mathcal{A}_2 - \mathcal{AE}_2, \mathsf{FO}_2(a) = \mathsf{IO}_2(a)$, and
        \item $\forall a \in \mathcal{AE}_2, \mathsf{FO}_2(a) \in 2^{\mathcal{P}} - \{\varnothing\} \land \mathsf{FO}_2(a) \neq \mathsf{IO}_2(a)$
    \end{itemize}
\end{itemize}
\end{definition}

In plain language, $\mathcal{GAE}$ represents the transfer of co-ownerships of one set of assets in one ledger in exchange for the transfer of co-ownerships of another set of assets in the other ledger, atomically. The standard atomic swap assumption of common parties across ledgers is represented by $\mathcal{P}$. Initial co-ownership of assets are represented by the definitions of $\mathsf{IO}_1$ and $\mathsf{IO}_2$ where each asset maps to a non-empty subset of $\mathcal{P}$. The definitions of $\mathsf{FO}_1$ and $\mathsf{FO}_2$ indicate that (i) the final co-owners of each asset being transferred are drawn from the full set of parties regardless of their initial ownership statuses, (ii) assets are not expunged from the ledger, and (iii) an asset's final co-ownership is different from its initial co-ownership. The function mappings remain unchanged for assets outside the swap lists $\mathcal{AE}_1$ and $\mathcal{AE}_2$. The range for each co-ownership function is the power set of $\mathcal{P}$, excluding the null set, indicating that each asset can have one or more co-owners drawn from $\mathcal{P}$. The null set is excluded because we wish to avoid handling assets in $\mathcal{GAE}$ that are created from scratch or destroyed in the course of a cross-ledger exchange. We limit our scenarios to assets exchanging hands because the incentive and trust models we consider in this paper only apply to those scenarios.

\begin{figure*}[ht]
    \centering
    \includegraphics[width=0.60\textwidth]{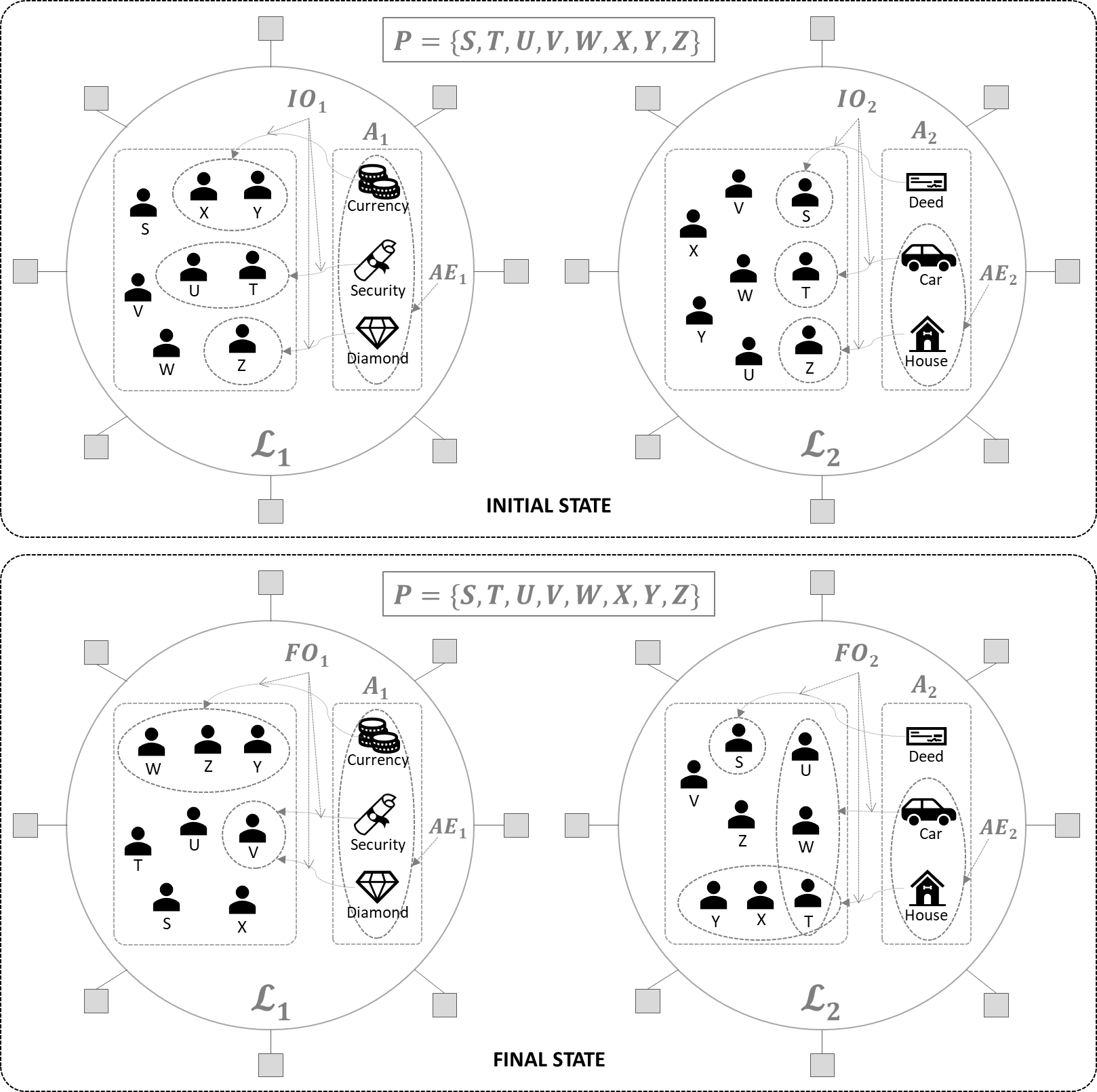}
    \caption{Generalized 2-Ledger Asset Exchange Example}
    \label{fig:complex-asset-exchange-example}
\end{figure*}

The example in Figure~\ref{fig:complex-asset-exchange-example} illustrates this definition and the tuple elements. From the definition, we can define partitions of the co-owners in order to reason about the exhaustiveness of the definition and the threat models they present, as follows for $i \in [2]$:
\begin{itemize}[leftmargin=*]
    \item $\mathsf{G}_i$ is the asset co-ownership giver function in $\mathcal{L}_i$, defined as $\mathsf{G}_i:\mathcal{AE}_i \to 2^\mathcal{P} - \{\varnothing\}$, where $\mathsf{G}_i(a) = \mathsf{IO}_i(a) - \mathsf{FO}_i(a)$
    \item $\mathsf{K}_i$ is the asset co-ownership keeper (sharer) function in $\mathcal{L}_i$, defined as $\mathsf{K}_i:\mathcal{AE}_i \to 2^\mathcal{P} - \{\varnothing\}$, where $\mathsf{K}_i(a) = \mathsf{IO}_i(a) \cap \mathsf{FO}_i(a)$
    \item $\mathsf{T}_i$ is the asset co-ownership taker function in $\mathcal{L}_i$, defined as $\mathsf{K}_i:\mathcal{AE}_i \to 2^\mathcal{P} - \{\varnothing\}$, where $\mathsf{K}_i(a) = \mathsf{FO}_i(a) - \mathsf{IO}_i(a)$
\end{itemize}
(\textit{Note}: as we mentioned earlier, the null set is excluded from the giver, keeper, and taker function ranges because we do not consider unattached assets that are created or destroyed in a $\mathcal{GAE}$ instance.)

As should be clear from these definitions, $\forall a \in \mathcal{AE}_1$, we have that $\mathsf{G}_1(a) \cup \mathsf{K}_1(a) \cup \mathsf{T}_1(a) = \mathsf{IO}_1(a) \cup \mathsf{FO}_1(a)$ and $\forall a \in \mathcal{AE}_2, \mathsf{G}_2(a) \cup \mathsf{K}_2(a) \cup \mathsf{T}_2(a) = \mathsf{IO}_2(a) \cup \mathsf{FO}_2(a)$.

\begin{table}[t]
\centering
\begin{tabular}{|c|c|c|c|c|c|} 
 \hline
  & \textbf{Ledger} & \textbf{Asset:Transfer} & \textbf{$\mathsf{G}$} & \textbf{$\mathsf{K}$} & \textbf{$\mathsf{T}$} \\
 \hline\hline
 \multirow{2}{*}{\textbf{ULT}} & $\mathcal{L}_1$ & $M: \{X\} \rightarrow \{Y\}$ & $\{X\}$ & $\varnothing$ & $\{Y\}$ \\
 \cline{2-6}
  & $\mathcal{L}_2$ & $N: \{Z\} \rightarrow \{W\}$ & $\{Z\}$ & $\varnothing$ & $\{W\}$ \\
 \hline\hline
 \multirow{2}{*}{\textbf{CLR}} & $\mathcal{L}_1$ & $M: \{X\} \rightarrow \{Y\}$ & $\{X\}$ & $\varnothing$ & $\{Y\}$ \\
 \cline{2-6}
  & $\mathcal{L}_2$ & $N: \{Z\} \rightarrow \{X\}$ & $\{Z\}$ & $\varnothing$ & $\{X\}$ \\
 \hline\hline
 \multirow{2}{*}{\textbf{CLS}} & $\mathcal{L}_1$ & $M: \{X\} \rightarrow \{Y\}$ & $\{X\}$ & $\varnothing$ & $\{Y\}$ \\
 \cline{2-6}
  & $\mathcal{L}_2$ & $N: \{Y\} \rightarrow \{X\}$ & $\{Y\}$ & $\varnothing$ & $\{X\}$ \\
 \hline
\end{tabular}
\caption{Co-Ownership Sets}
\label{table:gkt-coos}
\end{table}

Let us express cross-network transactions in $\mathcal{GAE}$ parlance, starting with the basic types discussed in Section~\ref{sec:basic-tx-types}. Table \ref{table:gkt-coos} shows the givers, keepers, and takers, of the transactions illustrated in Figure~\ref{fig:cross-network-tx-types}, each transaction involving the exchange of assets $M$ and $N$. We can spot a few conditionals from these tables:
\begin{itemize}[leftmargin=*]
    \item In the ULT pair, $(\mathsf{G}_1(M) \cup \mathsf{K}_1(M)) \cap \mathsf{T}_2(N) = \varnothing$ and $(\mathsf{G}_2(N) \cup \mathsf{K}_2(N)) \cap \mathsf{T}_1(M) = \varnothing$
    \item In the CLR, $(\mathsf{G}_1(M) \cup \mathsf{K}_1(M)) \cap \mathsf{T}_2(N) \neq \varnothing$ and $(\mathsf{G}_2(N) \cup \mathsf{K}_2(N)) \cap \mathsf{T}_1(M) = \varnothing$
    \item In the CLS, $(\mathsf{G}_1(M) \cup \mathsf{K}_1(M)) \cap \mathsf{T}_2(N) \neq \varnothing$ and $(\mathsf{G}_2(N) \cup \mathsf{K}_2(N)) \cap \mathsf{T}_1(M) \neq \varnothing$
\end{itemize}
We can therefore distinguish ULTs, CLRs, and CLSs using these conditionals. Table~\ref{table:gkt-complex} shows the co-ownership sets corresponding to the complex transaction of Figure~\ref{fig:complex-asset-exchange-example}. The above conditionals hold for the CLS transaction ($Currency$ for $House$), the CLR transaction ($Security$ for $Car$), and the ULT transaction ($Diamond$) are deconstructed earlier. If we merge all transactions in the two ledgers respectively and take the unions of the $\mathsf{G}$, $\mathsf{K}$, and $\mathsf{T}$ sets, we end up with the conditional above that corresponds to a CLS.

\begin{table}[t]
\centering
\begin{tabular}{|c|c|c|c|c|} 
 \hline
 \textbf{Ledger} & \textbf{Asset:Transfer} & \textbf{$\mathsf{G}$} & \textbf{$\mathsf{K}$} & \textbf{$\mathsf{T}$} \\
 \hline
 \multirow{3}{*}{$\mathcal{L}_1$} & $Currency: \{X,Y\} \rightarrow \{W,Y,Z\}$ & $\{X\}$ & $\{Y\}$ & $\{W,Z\}$ \\ \cline{2-5}
                      & $Security: \{T,U\} \rightarrow \{V\}$ & $\{T,U\}$ & $\varnothing$ & $\{V\}$ \\ \cline{2-5}
                      & $Diamond: \{Z\} \rightarrow \{V\}$ & $\{Z\}$ & $\varnothing$ & $\{V\}$ \\
 \hline
 \multirow{2}{*}{$\mathcal{L}_2$} & $Car: \{T\} \rightarrow \{T,U,W\}$ & $\varnothing$ & $\{T\}$ & $\{U,W\}$ \\ \cline{2-5}
                                & $House: \{Z\} \rightarrow \{T,X,Y\}$ & $\{Z\}$ & $\varnothing$ & $\{T,X,Y\}$ \\
 \hline
\end{tabular}
\caption{CLS Co-Ownership Sets}
\label{table:gkt-complex}
\end{table}

Now, with multiple assets and co-owners, consider the following sets of parties:
    \[
        \mathcal{S}_{12} = \left( \underset{a \in \mathcal{AE}_1}{\bigcup} \mathsf{G}_1(a)\; \cup \underset{a \in \mathcal{AE}_1}{\bigcup} \mathsf{K}_1(a) \right)\; \cap \underset{a \in \mathcal{AE}_2}{\bigcup} \mathsf{T}_2(a) 
    \]
    \[
        \mathcal{S}_{21} = \left( \underset{a \in \mathcal{AE}_2}{\bigcup} \mathsf{G}_2(a)\; \cup \underset{a \in \mathcal{AE}_2}{\bigcup} \mathsf{K}_2(a) \right)\; \cap \underset{a \in \mathcal{AE}_1}{\bigcup} \mathsf{T}_1(a)
    \]
We can thus extrapolate the conditionals to complex transactions with multiple assets and co-owners. $\mathcal{GAE}$ is a combination of 
\begin{itemize}[leftmargin=*]
    \item only gULTs if both $\mathcal{S}_{12}$ and $\mathcal{S}_{21}$ are empty.
    \item only gCLRs and gULTs if only one of $\mathcal{S}_{12}$ and $\mathcal{S}_{21}$ are empty. 
    \item all gCLSs, gCLRs and gULTs if neither of $\mathcal{S}_{12}$ and $\mathcal{S}_{21}$ is empty.
\end{itemize}


In plain language, if no giver or keeper gets anything in return for a transfer in another ledger, then the transactions are local and unconnected. If only a single group of co-owners get recompensed, the transactions will only involve cross-ledger replacements. And if there are two sets of co-owners get recompensed by each other, the transactions must involve cross-ledger swaps. Using the above test, we can verify that the examples in Figures~\ref{fig:asset-exchange-variation-1} and~\ref{fig:asset-exchange-variation-2} map to gCLSs and that the complex transaction in Figure~\ref{fig:complex-asset-exchange-example} contains gCLSs.

$\mathcal{GAE}$ therefore exhaustively covers all basic and complex transaction types as we have defined them. It also covers all possible cross-ledger exchanges in which atomicity is required. The above conditionals also tell us how to determine the combinations of basic transaction types within a $\mathcal{GAE}$ instance. 

\subsection{Threat model and Properties}
\label{sec:threat-model}
\noindent\textit{Threat model}: Let us identify the factors that pose threats to the integrity of a $\mathcal{GAE}$ instance. We assume that all participants in $\mathcal{P}$ are rational, and that they can arbitrarily deviate from a specified exchange protocol, collude with other participants, and attempt to maximize the value that they gain through the protocol. We also assume collective integrity of a network maintaining a shared ledger, i.e., rogue miners or block creators cannot override honest network members to finalize fraudulent or spam transactions.

\noindent\textit{Properties}: We desire the atomicity property from a protocol that accomplishes $\mathcal{GAE}$. Atomicity ensures for each party that follows the protocol that the assets it co-owns or would co-own either all move to the desired final state or all remain in the original state, even if other parties deviate from the protocol.
\begin{definition}[Atomicity in $\mathcal{GAE}$] \label{def:atomicity}
For every participant $P \in \mathcal{P}$, let $\mathcal{AE}_{1,P} = \{ a | P \in \left(\mathsf{IO}_1(a) \cup \mathsf{FO}_1(a) \right)\}$ and let $\mathcal{AE}_{2,P} = \{ a | P \in \left(\mathsf{IO}_2(a) \cup \mathsf{FO}_2(a) \right)\}$. A protocol for $\mathcal{GAE}$ is atomic if exactly one of the following holds: 
\begin{enumerate}[leftmargin=*]
    \item for all $a_1 \in \mathcal{AE}_{1,P}$, the owners of $a_1$ at the end of the protocol are according to $\mathsf{FO}_1(a_1)$ and for all $a_2 \in \mathcal{AE}_{2,P}$, the owners of $a_2$ at the end of the protocol are according to $\mathsf{FO}_2(a_2)$.
    \item for all $a_1 \in \mathcal{AE}_{1,P}$, the owners of $a_1$ at the end of the protocol are according to $\mathsf{IO}_1(a_1)$ and for all $a_2 \in \mathcal{AE}_{2,P}$, the owners of $a_2$ at the end of the protocol are according to $\mathsf{IO}_2(a_2)$.
\end{enumerate}
\end{definition}


\subsection{$\mathsf{HTLC}$ limitations}
\label{sec:htlc-limitations}
Let us categorize $\mathcal{GAE}$ instances from single-asset-per-ledger and single-co-owner-per-asset to multiples of these, and separately analyze the threats they face. In the basic CLR and CLS cases, and also in a case where a party keeps its co-ownership rather than gives it to another (see below), the conventional $\mathsf{HTLC}$ as described in Section~\ref{sec:introduction} can be used without it being susceptible to attack.
\begin{itemize}[leftmargin=*]
    \item $\mathcal{L}_1: M: \{X\} \rightarrow \{X,Y\}$
    \item $\mathcal{L}_2: N: \{Y\} \rightarrow \{X\}$
\end{itemize}

Now we will analyze all the permutations of $\mathcal{GAE}$ involving multiple co-owners and multiple assets in each ledger in below cases. Because $\mathsf{HTLC}$ is known to enforce atomicity for singly-owned assets, we will try to apply it to these scenarios to determine whether it satisfies the properties listed earlier.
\begin{enumerate}[leftmargin=*]
    \item \textit{Multiple co-owners locking a single asset in one ledger and claiming a single asset in another ledger}: This is the scenario in Figure~\ref{fig:asset-exchange-variation-1} and which we can represent as follows:
    \begin{itemize}
        \item $\mathcal{L}_1: M: \{X,W\} \rightarrow \{Y\}$
        \item $\mathcal{L}_2: N: \{Y\} \rightarrow \{X,W\}$
    \end{itemize}
     Conventional $\mathsf{HTLC}$ can be applied in two different ways: (i) $X$ and $W$ agree offline on a secret; either of them locks $M$ with its hash and signatures (i.e., consents) of both $X$ and $W$, following which $Y$ locks $N$ with the same hash; either $X$ or $W$ claims $N$ (on behalf of both parties), and then $Y$ claims $M$ (illustrated in Figure~\ref{fig:htlc-multi-signs} in Appendix \ref{appendix:HTLC-augmentations-and-attacks}), (ii) $X$ and $W$ independently lock $M$ in either sequence using two different secrets (the contract deems $M$ to be locked only when both $X$ and $W$ have submitted their respective transactions), following which $Y$ locks $N$ with both the hashes; $X$ and $W$ submit claims to $N$ (by revealing their respective secrets in any order), following which $Y$ claims $M$ using both the revealed secrets (Figure~\ref{fig:htlc-multi-secrets} in Appendix \ref{appendix:HTLC-augmentations-and-attacks}).
    
    \textit{Threats/Attacks}: Both approaches are vulnerable to collusion between $Y$ and one of $X$ and $W$ to cheat the other: (i) after $M$ is locked, either of its co-owners (let's say $W$) may strike a deal with $Y$ to give $M$ away (using the same hash lock) and $Y$ in turn will lock $N$ only for that party (in this case, $W$) rather than jointly to $X$ and $W$; $W$ may also provide some other incentive to $Y$ to encourage it to collude against $X$; since $X$ has already locked away its co-ownership of $M$ using a secret that $W$ also possesses, it is left at a disadvantage (illustrated in Figure \ref{fig:attack-on-htlc-multi-signs} in Appendix \ref{appendix:HTLC-augmentations-and-attacks}) (ii) if $X$ submits its claim in $\mathcal{L}_2$ first by revealing its secret, $W$ may avoid revealing its secret and submitting a claim, following which $Y$ may allow its earlier lock to lapse and instead lock $N$ again only in favor of $W$; having revealed its secret, $X$ is now at a disadvantage as $Y$ can claim its share of $M$ but $W$ now will get all of $N$ instead of having to co-own it with $X$ (illustrated in Figure \ref{fig:attack-on-htlc-multi-secrets} in Appendix \ref{appendix:HTLC-augmentations-and-attacks}). In this protocol, the party that acts first (i.e., submits a claim) stands at a disadvantage as its co-owner may collude with its counterparty(ies) to rob it of its rightful share of an asset.
    
    Incidentally, if $Y$ locks $N$ first, these attacks will not be possible, but if $Y$ is replaced with multiple co-owners, then these threats will still apply regardless of the lock sequence. 
    \item \textit{Co-owners locking multiple assets in one ledger and claiming a single asset in another ledger}: This is the scenario in Figure~\ref{fig:asset-exchange-variation-2}:
    \begin{itemize}
        \item $\mathcal{L}_1: M: \{X\} \rightarrow \{Y\}$
        \item $\mathcal{L}_1: R: \{W\} \rightarrow \{Y\}$
        \item $\mathcal{L}_2: N: \{Y\} \rightarrow \{X,W\}$
    \end{itemize}
    Conventional $\mathsf{HTLC}$ can be applied in the same two ways as in case (1) and suffers from the same collusion possibilities. The only difference is that $Y$ gains two assets instead of one if it colludes with either $X$ or $W$. If $Y$ were to lock $N$ first, it could simply lock in favor of either $X$ or $W$ (whoever the colluding party is). This procedure does not present any interesting features, because $Y$ has the power of unilateral action regardless of how the other parties behave. The same threat possibilities apply if $Y$ is replaced with multiple co-owners.
    \item \textit{Multiple co-owners locking a single asset in one ledger and claiming multiple assets in another ledger}:
    \begin{itemize}
        \item $\mathcal{L}_1: M: \{X,W\} \rightarrow \{Y\}$
        \item $\mathcal{L}_2: R: \{Y\} \rightarrow \{X\}$
        \item $\mathcal{L}_2: N: \{Y\} \rightarrow \{W\}$
    \end{itemize}
    If conventional $\mathsf{HTLC}$ is applied with $M$ being locked first, it will suffer from the same threats as in cases (1) and (2). But if $Y$ were to lock both $R$ and $N$ in $\mathcal{L}_2$ first, then conventional $\mathsf{HTLC}$ will work without posing any new threats.
    
    If $Y$ is replaced with multiple co-owners, say $\{Y,Z\}$, the attacks described in case (1) will be possible (either $Y$ or $Z$ colluding with $X$ and $W$).
    \item \textit{Co-owners locking multiple assets in one ledger and claiming multiple assets in another ledger}:
    \begin{itemize}
        \item $\mathcal{L}_1: M: \{X\} \rightarrow \{Y\}$
        \item $\mathcal{L}_1: R: \{W\} \rightarrow \{Y\}$
        \item $\mathcal{L}_2: N: \{Y\} \rightarrow \{X\}$
        \item $\mathcal{L}_2: P: \{Y\} \rightarrow \{W\}$
    \end{itemize}
     If conventional $\mathsf{HTLC}$ is applied with $M$ and $R$ being locked first as in case (2), it will face the same threats. If $Y$ were to lock $N$ and $P$ first, it could simply lock one of those assets in favor of either $X$ or $W$ (whoever the colluding party is). On the other hand, if $Y$ locked both its assets first and then waited for both $X$ and $W$ to lock theirs, then $\mathsf{HTLC}$ will fulfil the exchange. But if $Y$ were replaced with multiple co-owners, the same threat possibilities apply as in case (1).
\end{enumerate}
By induction, the above breakup of cases and the threats identified can be extrapolated to a generic $\mathcal{GAE}$ instance. The threat faced by conventional $\mathsf{HTLC}$ when applied to $\mathcal{GAE}$ is that there is no mechanism to enforce a \textit{joint lock and claim} over more than one asset by more than one co-owner. We must therefore (i) ensure that parties can jointly lock one or more assets without revealing the secret to any of them in clear, and (ii) ensure that parties jointly locking one or more assets for transfer in one ledger cannot unilaterally make claims in the other ledger. We will address these challenges by creating primitives for these operations and providing an augmented $\mathsf{HTLC}$ protocol in the next section.

\section{Solution: Multi-Party Hash Time Locked Contract (MPHTLC)}
\label{sec:solution}

We present an augmented form of $\mathsf{HTLC}$, called $\mathsf{MPHTLC}$, to solve $\mathcal{GAE}$.
The basic capabilities required of a ledger (or a smart contract system over it) is that it supports hash and time locks on assets. Therefore, any ledger that is capable of implementing conventional $\mathsf{HTLC}$~\cite{NolanSwap13} will also be able to implement $\mathsf{MPHTLC}$.
We also make the practical assumption that both participating ledgers have known (or predictable) upper and lower bounds for transaction, or block, confirmation times. As a consequence, we can safely assume that no adversary in our threat model can arbitrarily speed up or slow down either of the ledgers. This ensures predictability and a priori selection of asset time lock durations, preventing the counterparties from racing each other or denying each other adequate time to mount their claims.
In addition, we will utilize \textit{secure multi-party computation} (MPC) protocols~\cite{Yao82,GMW87,fairMPCCCS17}, to enforce the ability to jointly lock and claim assets, which we identified as a key requirement to handle the threats.

\subsection{$\mathsf{MPHTLC}$ protocol}
\label{sec:mphtlc-protocol}
In a $\mathcal{GAE}$ instance, let us define $\mathcal{O}_1$ as the set of initial co-owners of assets $\mathcal{AE}_1$ in $\mathcal{L}_1$ (givers and keepers, as defined in Section~\ref{sec:modeling}) as
    \[
        \mathcal{O}_1 = \underset{a \in \mathcal{AE}_1}{\bigcup} \mathsf{IO}_1(a) = \left(\underset{a \in \mathcal{AE}_1}{\bigcup} \mathsf{G}_1(a)\; \cup \underset{a \in \mathcal{AE}_1}{\bigcup} \mathsf{K}_1(a)\right)
    \]
Let the cardinality of $\mathcal{O}_1$, or $|\mathcal{O}_1|$, be $n$. Here are the protocol steps:


\begin{enumerate}[leftmargin=*]
    \item Let $x_1,x_2,\ldots,x_n$ be the secrets chosen by the $n$ parties of $\mathcal{O}_1$ respectively in $\mathcal{L}_1$. They compute a hash $H = F_1(x_1,x_2,\ldots,x_n)$ where $F_1$ is an MPC protocol (see Section \ref{sec:mpc-instantiation}) among the $n$ parties. Note: no information about the secrets $x_1,x_2,\ldots,x_n$ is revealed during the computation other than the output $H$ itself.
    \item Each asset $a \in \mathcal{AE}_1$ is locked by one of its initial co-owners ($o \in \mathsf{IO}_1(a)$) using hash $H$ in $\mathcal{L}_1$ with expiration duration $T$ in favor of the parties in $\mathsf{FO}_1(a)$. All initial co-owners of $a$ in $\mathsf{IO}_1(a)$ are required to sign the lock transaction before it is submitted to the chain. It is up to the ledger/contract to determine if each asset is to be locked using a separate transaction or if all assets are to be locked together in a single transaction. The contract(s) validate the signatures from all the co-owners of each asset and if valid, locks the assets pending either the supply of the secret pre-image corresponding to $H$ or the passing of $T$ time units. 
    \item Each asset $a \in \mathcal{AE}_2$ is locked by one of its initial co-owners ($o \in \mathsf{IO}_2(a)$) after verifying that all the locks in Step (2) have been committed in favor of the expected final owners with appropriate time locks in $\mathcal{L}_1$.
    The locks in $\mathcal{L}_2$ are created using same hash $H$ from $\mathcal{L}_1$ with an expiration duration $T/2$ in favor of the parties in $\mathsf{FO}_2(a)$, pending either the supply of the secret pre-image corresponding to $H$ or the passing of $T/2$ time units. All initial co-owners of $a$ in $\mathsf{IO}_2(a)$ are required to sign the lock transaction before it is submitted to the chain. Note that hash $H$ can be obtained from $\mathcal{L}_1$, because $\mathsf{IO}_2(a) \subseteq \mathcal{P}$ and all members of $\mathcal{P}$ have read access to both ledgers. 
    \item The $n$ parties of $\mathcal{O}_1$ in $\mathcal{L}_1$ first verify that all the locks in Steps (2) and (3) have been committed in favor of the expected final owners with appropriate time locks in $\mathcal{L}_1$ and $\mathcal{L}_2$ respectively. If these checks are successful, the $n$ parties compute $x=F_2(x_1,x_2,\ldots,x_n)$ such that $Hash(x) = H$, using their respective secrets chosen in Step $1$ as inputs. Here, $F_2$ is an MPC protocol (described in Section \ref{sec:mpc-instantiation}) between the $n$ parties.
    \item Every asset $a \in \mathcal{AE}_2$ can now be claimed by its final owners $\mathsf{FO}_2(a)$ in $\mathcal{L}_2$ if the secret $x$ is revealed before time duration $T/2$ elapses. If the transfer of $a$ is part of a gCLS or a gCLR, i.e., if the intersection of $\mathcal{O}_1$ and $\mathsf{FO}_2(a)$ is not $\varnothing$, there will be at least one party with an incentive to reveal the secret. If the intersection is $\varnothing$, then the transfer of $a$ is a gULT with respect to the transfers in $\mathcal{L}_1$; as described in Section~\ref{sec:basic-tx-types}, this will require an off-chain sharing of the secret among parties if and only if this gULT must be conducted atomically with transfers in $\mathcal{L}_1$. All assets in $\mathcal{AE}_2$ get transferred to the final co-owners as defined by the $\mathsf{FO}_2$ function once any party reveals $x$ in $\mathcal{L}_2$ within $T/2$. At this stage, the revelation of the secret benefits all claiming parties equally, and gives no advantage to a subset. If nobody supplies $x$ in $\mathcal{L}_2$ within $T/2$, the co-owners remain as defined by $\mathsf{IO}_2$. All transfers occur or no transfer occurs, thereby satisfying the cross-ledger atomicity property. Lastly, depending on how the assets were locked, the claim should be submitted via a single transaction for all locked assets or using one transaction per locked asset. 
    \item Every asset $a \in \mathcal{AE}_1$ can now be claimed by its final owners $\mathsf{FO}_1(a)$ in $\mathcal{L}_1$ by supplying the secret $x$ within duration $T$. $x$ is already public in $\mathcal{L}_2$ and therefore known to every party in $\mathcal{P}$. Every asset in $\mathcal{AE}_1$ gets transferred to its final co-owners as defined by $\mathsf{FO}_1$ once any party reveals $x$ in $\mathcal{L}_1$ within $T$. If nobody supplies $x$ in $\mathcal{L}_1$ (which would be irrational on their part), then the co-owners remain as defined by $\mathsf{IO}_1$. 
\end{enumerate}
\textit{Timing Considerations}: If each asset has a  separate lock and claim transaction, there is a potential hazard that must be managed as follows. In Step $5$, there may be an asset $a \in \mathcal{AE}_2$ whose claimers don't overlap with $\mathcal{O}_1$, i.e., $\mathsf{T}_2(a) \cup \mathcal{O}_1 = \varnothing$. These claimers are dependent on some party in $\mathcal{O}_1$ revealing $x$. But if parties in $\mathcal{O}_1$ compute MPC $F_2$ to reveal $x$ close to the timeout $T/2$, these non-overlapping claimers may lose the chance to claim their assets. Therefore, the time locks on such assets should have extra tolerance. And because we can determine these assets a priori, by inspecting the $\mathcal{GAE}$ tuple, the contracts can enforce this. A suggestive guideline would be to use $T/4$ as the timeout for every asset in $\mathcal{L}_2$ that has at least one claimant belonging to $\mathcal{O}_1$, and $T/2$ for the other assets in $\mathcal{L}_2$. The timeout for every asset in $\mathcal{L}_1$ remains $T$ as $x$, once revealed in $\mathcal{L}_2$, is immediately available for claiming in $\mathcal{L}_1$.


\subsection{Selection of the MPC functions $F_1$ and $F_2$}
\label{sec:mpc-instantiation}
The $n$ parties output $H = Hash\left(g(x_1,x_2,\ldots,x_n)\right)$ by participating in an MPC protocol $F_1$ with $(x_1,x_2,\ldots,x_n)$ as their respective inputs, where $g$ can be any function with the following properties:
\begin{enumerate}[leftmargin=*]
    \item The output of $g$ contains sufficient entropy for input from each party, such that inputs of any $n-1$ parties cannot be used to guess the output with non-negligible probability.
    \item The output of function $g$ need not hide information about the inputs $x_1,x_2,\ldots,x_n$. This is what differentiates $g$ from, say, a secret sharing protocol.
    \item The one-wayness of $H$ ought to be retained when hashing $g(x_1,x_2,\ldots,x_n)$ instead of $x_1,x_2,\ldots,x_n$. \textit{Note}: the security of $\mathsf{HTLC}$ assumes one-wayness of $H$ on randomly chosen inputs.
\end{enumerate}
In our instantiation of $\mathsf{MPHTLC}$, $g$ is an identity function which outputs $x_1, x_2, \ldots, x_n$, but any $g$ satisfying the above properties can be used. Any maliciously secure MPC protocol with fairness can be used for $F_1$, where the malicious security protects against participants arbitrarily deviating from the protocol and the fairness property guarantees that either all the parties get the output or none of them do. 
The fairness property ensures that the same output hash is obtained by all the parties in $\mathcal{O}_1$ and is a valid hash according to the inputs $x_1, x_2, \ldots, x_n$. Note again that no information is revealed about the secrets $x_1,x_2,\ldots,x_n$ during the computation of $Hash\left(g(x_1,x_2,\ldots,x_n)\right) = H$, other than the final output $H$. The above mentioned properties of $g$ do not affect the confidentiality provided by the MPC protocol.

The MPC protocol $F_2$ among the $n$ parties of $\mathcal{O}_1$ in $\mathcal{L}_1$ (used during the claim phase) involves computing $x = g(x_1,x_2,\ldots,x_n)$ using their respective inputs $x_1, x_2, \ldots, x_n$ chosen in Step $1$. Note that $Hash(x) = H$.
We use an MPC protocol with fairness and hence either all the parties in $\mathcal{O}_1$ obtain $x$ or none of them do.\\

\noindent{\bf A discussion on Fair MPC protocols.}
Fairness in MPC is extensively studied in the cryptography literature, and there are different possibilities of instantiating a fair MPC based on the requirements of $\mathsf{MPHTLC}$. It is possible to achieve fairness with just point-to-point secure communication between each pair of individual parties if less than one-third of the parties are malicious \cite{GMW87, BGW88}, and with the use of a broadcast network to protect up to $\lceil n/2 \rceil - 1$ parties being malicious \cite{GMW87, RB89}. 
If the goal is to protect against a majority of the parties being malicious, it is proven that the standard definition of fairness is impossible to achieve this without additional assumptions \cite{Cle86}, and this impossibility holds even in the presence of a trusted hardware without trusted clocks \cite{PST17}. There are different relaxed definitions of fairness proposed in the literature that protect against a dishonest majority of parties \cite{ASW98, Blu83, GL05, GMPY06}. But if a use of $\mathsf{MPHTLC}$ demands the standard definition of fairness against a dishonest majority of parties, there are protocols that use public blockchains, financial penalties and sometimes additionally trusted hardware like Intel SGX (to make the protocol more efficient) to enable this \cite{BK14, CGJ+17, KZZ16, PST17, BDD20}.

The fairness property is sufficient if all the assets are locked in a single transaction. But in the variant that assets are locked separately across multiple transactions, there is an additional hazard that we need to address. Fairness ensures that either all the parties get the output or none of them do, but it does not specify any time bound on when the parties get the output if they do. There could exist fair MPC protocols where one of the parties delays the output ``release'' to whenever they want to. This is a concern when the assets are individually locked since a party can delay the output release to just before the timeout $T/2$, and hence, due to network delays, there will invariably be some parties at a disadvantage. Though all the parties will get the output, some of them might get it before the timelock expires and some might get it after. To protect against this, we need a ``timeliness'' property in the fair MPC protocol which ensures that, if the parties were to get the output, there is a pre-specified time bound before which all of them should get it, and after which none of them should. We note that it is possible to augment the protocol in \cite{CGJ+17} based on secure hardware with this property, additionally assuming the existence of a trusted clock inside the trusted hardware \cite{inteltrustedclock, towncrier}. Informally, in \cite{CGJ+17}, there is a release token posted on a bulletin board (instantiated using a public blockchain) to initiate the release of the MPC output. A proof of posting this token is then verified by a program installed and attested in the secure hardware before it returns the MPC output. Now, this program additionally needs to check whether the release token is posted on the bulletin board before a certain time bound. We defer a detailed discussion on fair MPC protocols with the timeliness property to future work.

\subsection{Instantiations}
\label{sec:use-cases-security}

We now instantiate the $\mathsf{MPHTLC}$ protocol for a couple of scenarios, that also provides an intuition on how $\mathsf{MPHTLC}$ avoids the attacks identified as possible with conventional $\mathsf{HTLC}$ in Section~\ref{sec:threat-model}. Across all the cases, we discovered that collusion attacks are possible when
\begin{itemize}[leftmargin=*]
    \item multiple co-owners must lock an asset first before making a joint claim for an asset in another ledger, as described in case (1) in Section~\ref{sec:threat-model} (also see Figure~\ref{fig:asset-exchange-variation-1})
    \item co-owners of different assets must lock their respective assets first before making a joint claim for an asset in another ledger, as described in case (2) in Section~\ref{sec:threat-model} (also see Figure~\ref{fig:asset-exchange-variation-2})
\end{itemize}

\begin{figure*}[ht]
    \centering
    \includegraphics[width=0.65\textwidth]{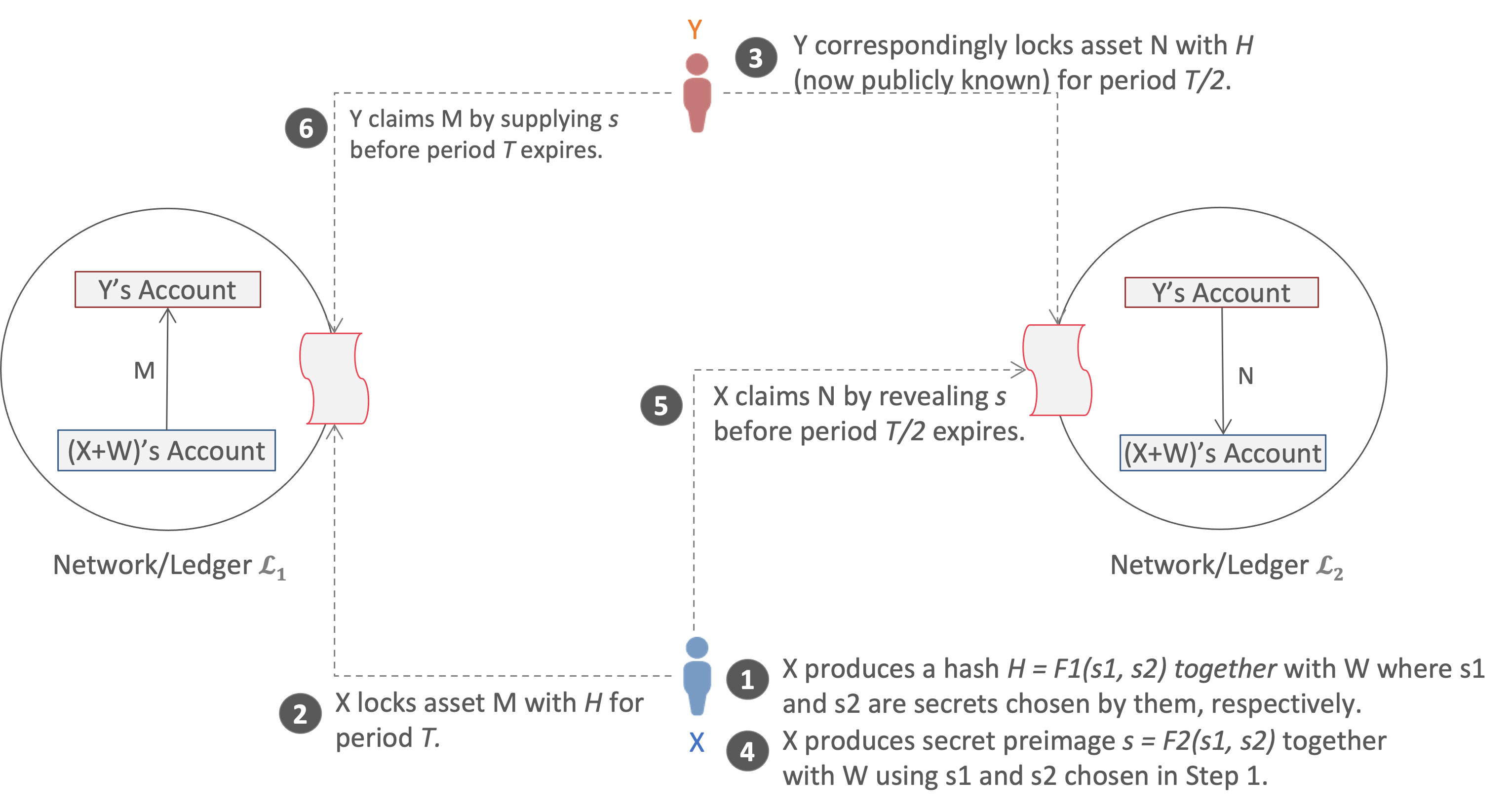}
    \caption{Multi-Party Hash Time Locked Contract ($\mathsf{MPHTLC}$) Protocol}
    \label{fig:mphtlc-caas}
\end{figure*}
$\mathsf{MPHTLC}$ applied to case (1) goes as follows (also see Figure~\ref{fig:mphtlc-caas}).
\begin{enumerate}[leftmargin=*]
    \item Let $s_1$ and $s_2$ be the secrets chosen by $X$ and $W$ respectively. They compute a hash $H=F_1(s_1,s_2)$ where $F_1$ is a MPC protocol instance agreed by both of them.
    \item Party $X$ locks the asset $M$ using $H$ in $\mathcal{L}_1$ till time $T$ with the signature of the party $W$ for party $Y$, or vice versa.
    \item Party $Y$ locks the asset $N$ using $H$ in $\mathcal{L}_2$ till time $T/2$ for both the parties $X$ and $W$.
    \item Parties $X$ and $W$ in $\mathcal{L}_1$ compute $s=F_2(s_1,s_2)$ using their respective secrets chosen in Step $1$ above where $F_2$ is a another MPC protocol instance chosen by them such that $Hash(s) = H$.
    \item Party $X$ (or $W$) submits a claim on asset $N$ in $\mathcal{L}_2$ by revealing the secret $s$ before time $T/2$ elapses.
    \item Party $Y$ submits a claim on asset $M$ in $\mathcal{L}_1$ by using the secret $s$ (obtained from ledger $\mathcal{L}_2$) before time $T$ elapses.
\end{enumerate}
This protocol avoids the two hazards identified in Section~\ref{sec:threat-model}, at lock time and at claim time, as follows:
\begin{itemize}[leftmargin=*]
    \item Neither $X$ nor $W$ knows the secret at the end of Step 2, and hence neither can collude with $Y$ to get the latter to lock $N$ only in favor of one of them. If $Y$ chooses not to execute Step 3, the protocol will end with no change in either ledger's state.
    \item There is exactly one secret, which is only revealed in Step 4, so $X$ nor $W$ will be left at a disadvantage. If either of them reveals $s$ in that step, both will collectively co-own $N$. If neither reveals $s$, the protocol will end with no change in either ledger's state.
\end{itemize}



$\mathsf{MPHTLC}$ applied to case (2) goes as follows (also see Figure~\ref{fig:mphtlc-maas} in Appendix \ref{appendix:MPHTLC-workflow-instance}).
\begin{enumerate}[leftmargin=*]
    \item Let $s_1$ and $s_2$ be the secrets chosen by $X$ and $W$ respectively. They compute a hash $H=F_1(s_1,s_2)$ where $F_1$ is an MPC protocol instance agreed by both of them.
    \item Party $X$ locks asset $M$ using $H$ in $\mathcal{L}_1$ for time $T$ for party $Y$. Similarly, $W$ locks asset $R$ using $H$ in $\mathcal{L}_1$ for time $T$ for $Y$.
    \item Party $Y$ locks asset $N$ using $H$ in $\mathcal{L}_2$ for time $T/2$ for both the parties $X$ and $W$.
    \item Parties $X$ and $W$ in $\mathcal{L}_1$ compute $s=F_2(s_1,s_2)$ using their respective secrets chosen in Step $1$ above where $F_2$ is an another MPC protocol instance chosen by them such that $Hash(s) = H$.
    \item Party $X$ (or $W$) submits a claim on asset $N$ in $\mathcal{L}_2$ by revealing the secret $s$ before time $T/2$ elapses.
    \item Party $Y$ submits a claim on assets $M$ and $X$ in $\mathcal{L}_1$ by using the secret $s$ (obtained from ledger $\mathcal{L}_2$) before time $T$ elapses.
\end{enumerate}
This protocol avoids the same two hazards in the same ways as discussed for the previous use case.

\section{Analysis}
\label{sec:analysis}
We now prove the atomicity property of our $\mathsf{MPHTLC}$ protocol in our threat model that all parties are rational and the adversarial parties can arbitrarily deviate from the protocol. 

We state some basic assumptions before we prove atomicity:
\begin{itemize}[leftmargin=*]
    \item At least one party in $\mathcal{O}_1$ has incentive to claim a locked asset in $\mathcal{L}_2$.
    \item For each asset locked in $\mathcal{L}_1$, there exists at least one party in $\mathcal{P}$ that has an incentive to claim it.
    \item The time lock duration can be determined by the parties based on the environmental considerations to ensure that the claimers are left with sufficient time.
    \item The completion time of the fair MPC protocol is predictable.
\end{itemize}

\begin{claim}
$\mathsf{MPHTLC}$ is atomic according to Definition \ref{def:atomicity}.
\end{claim}
\begin{proof}
We will start by arguing atomicity for the variant of $\mathsf{MPHTLC}$ where there is a single transaction in $\mathcal{L}_1$ (respectively $\mathcal{L}_2$) which locks all the assets in $\mathcal{AE}_1$ (respectively $\mathcal{AE}_2$). The honest parties (that follow the protocol) in $\mathcal{O}_1$ proceed with the MPC $F_2$ only if the contracts are appropriate in $\mathcal{L}_1$ and $\mathcal{L}_2$, and if they proceed, the fairness property of the MPC protocol ensures that all or none of the parties in $\mathcal{O}_1$ obtain the pre-image. If none of them obtain the pre-image, then all assets in $\mathcal{AE}_2$ and later the ones in $\mathcal{AE}_1$ revert to their initial owners. And if all the parties obtain the pre-image, since at least one of them has the incentive to do the claim in $\mathcal{L}_2$, the assets in $\mathcal{AE}_2$ would go to the final owners according to $\mathsf{FO}_2$. And on seeing the pre-image used, a party that has an incentive to claim a locked asset in $\mathcal{L}_1$ (we assumed there exists at least one) does so. The protocol allows sufficient time for this party to succeed with the claim in $\mathcal{L}_1$, and hence the assets in $\mathcal{AE}_1$ would also go their final owners as prescribed by $\mathsf{FO}_1$.

In the variant where each asset is locked and claimed via separate transactions, the extra time available for parties not in $\mathcal{O}_1$ facilitates their asset claims even if no party in $\mathcal{O}_1$ submits a claim in $\mathcal{L}_2$ until very close to their timeout. The rest of the argument follows, with the help of our assumption that there exists at least one party in $\mathcal{P}$ to have the incentive to claim each locked asset in $\mathcal{L}_1$.
\end{proof}

\begin{claim}
$\mathsf{MPHTLC}$ results in a Nash equilibrium for parties engaged in a  set of asset swaps (gCLSs). 
\end{claim}

We provide an informal proof of this claim using a high-level game-theoretic argument, based on a proof in Belotti et al.~\cite{BMPSICDCS20} that conventional $\mathsf{HTLC}$ results in Nash equilibrium. We limit $\mathcal{GAE}$ scenarios under consideration to those involving only gCLSs (i.e., all parties are involved in a set of swaps across 2 ledgers, like the scenarios in Figures~\ref{fig:asset-exchange-variation-1} and~\ref{fig:asset-exchange-variation-2}).
We argue that if none of the target co-owners of an asset $a$ (i.e., $\mathsf{FO}_2(a)$) bother to claim it in Step $5$ (after completing Step $4$ and revealing the secret), then it is guaranteed that at least one or more of these co-owners always get cheated by one or more other co-owners, who can collude with current co-owners of $a$ (i.e., $\mathsf{IO}_2(a)$) in $\mathcal{L}_2$. (This is the hazard we identified in Section~\ref{sec:htlc-limitations} and avoided in $\mathsf{MPHTLC}$.)

We informally prove our argument with reference to the gCLS illustrated in Figure~\ref{fig:asset-exchange-variation-1}, where multiple co-owners of a single asset in one ledger are attempting to swap another single asset in the other ledger with a single owner. Let us follow the $\mathsf{MPHTLC}$ protocol applied to this scenario (see Section~\ref{sec:use-cases-security}). Let's assume the protocol has proceeded as expected until the end of Step 4. Now, if neither $X$ nor $W$ bothers to claim $N$ in $\mathcal{L}_2$ before time $T/2$ elapses, $Y$'s lock expires. Now, one of the co-owners of $M$ (say $W$) can collude with $Y$ to obtain $N$ all for itself (excluding $X$) via a hash lock using the same $H$. Then it proceeds to reveal secret $s$, and claims $N$. $Y$ in turn can claim $M$ in $\mathcal{L}_1$ using $s$ before $T$ elapses.

But note that this fraud on the part of $W$ requires it to race $X$ to collude with $Y$, and there is no guarantee that it will win. The loser of this race stands to lose all of $N$ instead of a co-ownership of $N$, which it was guaranteed to get had it followed protocol (i.e., Step 5). Therefore, it is in the interest of both $W$ and $X$ to reveal the secret in Step 4 rather than let $T/2$ elapse without claiming $N$. As we assumed that all parties in a $\mathcal{GAE}$ are rational, none of them would pick the alternate strategy described above. Speaking generally, at least one member of $\mathcal{O}_1$ is guaranteed to reveal the secret within $T/2$. Hence, following the $\mathsf{MPHTLC}$ protocol results in the best outcome for the parties involved (an equilibrium), and there is no incentive for parties to deviate from it. 





\section{Implementation}
\label{sec:implementation}





Any DLT that can support $\mathsf{HTLC}$ can also support $\mathsf{MPHTLC}$ through simple changes in the core logic of the locking and claiming modules (DLT's smart contract). For a DLT to support $\mathsf{MPHTLC}$ (or plain $\mathsf{HTLC}$), it needs to support the locking (or freezing) of state for a finite time period (a \textit{time lock}).

A typical DLT application model has 2 layers: the Contracts layer has business logic in contracts which run on peers and update state through consensus, whereas the Applications layer consists of client applications which invoke transactions on contracts. Any asset exchange protocol ($\mathsf{HTLC}$ or $\mathsf{MPHTLC}$) requires implementation in both layers to protect asset integrity against Byzantine failures; just an Applications layer implementation without decentralized trust (consensus) guarantees cannot protect asset integrity. Core $\mathsf{MPHTLC}$ capabilities of locking, claiming, and signature verifications can be built as smart contracts while client layer apps can submit lock and claim transaction requests to these contracts.

We will show how an existing atomic swap capability built on $\mathsf{HTLC}$ can be augmented to support $\mathsf{MPHTLC}$ between permissioned DLTs like Hyperledger Fabric and Corda. From several interoperation framework candidates (Polkadot \cite{Polkadot16}, Cosmos \cite{Cosmos16}, Cactus \cite{Cactus20} and Weaver \cite{Weaver21}), we picked Hyperledger Labs Weaver~\cite{Weaver21}, which supports $\mathsf{HTLC}$ and does not require changes to Fabric or Corda. (\textit{Note}: our protocol can be implemented on a permissionless network too, as long as the network supports conventional $\mathsf{HTLC}$.)

\begin{figure}[ht]
    \centering
    \includegraphics[width=0.99\columnwidth]{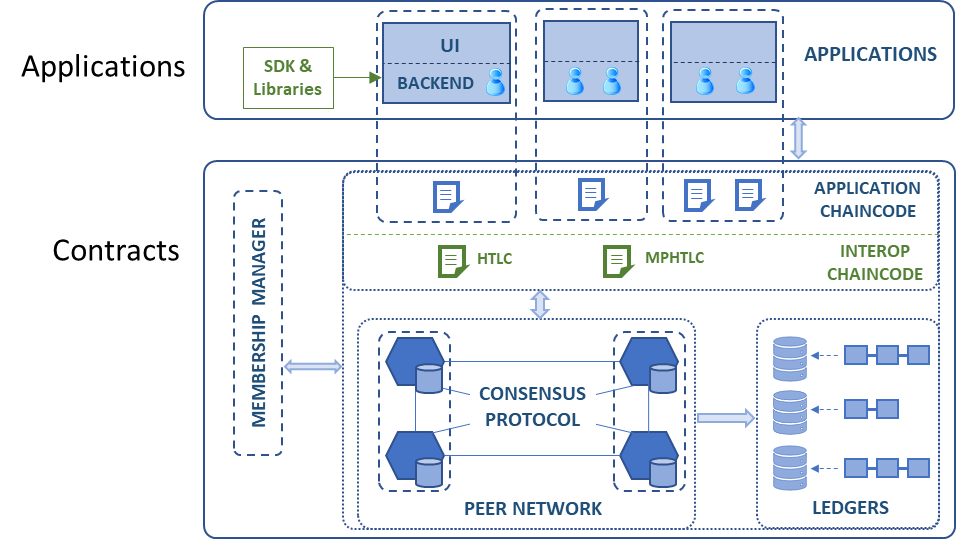}
    \caption{Fabric Network and Decentralized App Architecture}
    \label{fig:implementation-fabric}
\end{figure}

For Fabric, Weaver supports $\mathsf{HTLC}$ through: (i) a special \textit{chaincode} (contract) called the \textit{Interop Chaincode} that contains hash-time locks, and asset claim operations, and (ii) client library (API/SDK) to trigger lock and claim transactions from the Applications layer (see Figure  \ref{fig:implementation-fabric}). Time locks are enforced in contract logic by comparing the timestamp presented in a claim transaction to a peer's local time; the assumption is that peers' local times are approximately in sync and therefore a consensus on expiration can be reached. Any asset management application's chaincode needs to be modified only to trigger locks and claims by invoking the \textit{Interop Chaincode}. The client application can then submit asset lock and claim instructions to its chaincode using the Weaver API and libraries, and expose hashes and secrets after revelation via API functions.

We implemented our protocol by creating a fork\footnote{\href{https://github.com/mphtlc/weaver-dlt-interoperability}{github.com/mphtlc/weaver-dlt-interoperability}} of open source Weaver repository. To support $\mathsf{MPHTLC}$ in Fabric, we implemented the capabilities described in Section~\ref{sec:mphtlc-protocol} directly in the Weaver \textit{Interop Chaincode} and client SDK. Application developers will need to change very little (just additional API calls in both layers) apart from enforcing the properties of asset co-ownerships.



In Corda, \textit{CorDapps} deployed on network nodes scope business logic in the Contracts layer and \textit{flows} (in the Applications layer) (see Figure \ref{fig:implementation-corda} in Appendix \ref{weaver-arch-corda}). To support $\mathsf{HTLC}$, Weaver offers a special \textit{Interop CorDapp} that contains features similar to those described earlier for Fabric (locks, claims, and time lock enforcement). Any CorDapp that manages assets must, as in the Fabric case, be augmented to exercise the \textit{Interop CorDapp} (contract features and client libraries). To support $\mathsf{MPHTLC}$ in Corda using Weaver, we modified the \textit{Interop CorDapp} to create a flow so that multiple co-owners (instead of a single one) in a $\mathcal{GAE}$ instance sign the lock transactions (Steps 2 and 3 in Section~\ref{sec:mphtlc-protocol}), while multiple co-owners (instead of a single one) sign the claims (Steps 5 and 6 in Section~\ref{sec:mphtlc-protocol}). As an aside, multiple co-owners can reclaim/unlock an asset if it is unclaimed before timeout. A \textit{CorDapp} developer need only insert calls to these bulk lock/claim/unlock functions in his code.
( See Appendices \ref{sec:appendix1} and \ref{sec:appendix2} for API details for both DLTs.)

\textit{Experimental Evaluation}: We evaluated our Weaver implementation of  $\mathcal{GAE}$ by creating a sample application for co-owned assets in both Corda and Fabric. We ran our experiments on a Virtual Machine with 8-cores of Intel(R) Xeon(R) Gold 6140 CPU @ 2.30GHz, 16GB RAM, and 64-bit Ubuntu 18.04 as the OS. Table \ref{tab:mphtlc-results-latency} reports the latency for locking an asset, querying a locked asset state, claim, and unlock. Latency is computed as the time taken by the respective API (SDK) calls to complete execution. 2nd (Fabric 2P) and 3rd (Corda 2P) columns in Table \ref{tab:mphtlc-results-latency} report latencies for asset exchange operations for two parties co-owning one shared asset before and after the asset exchange.
We also observe that the time required for the API execution is proportional to the number of parties involved in each of these transactions (lock/claim/unlock) since we need to collect the consent from all these parties. We expect the time to run the MPC tasks $F_1$ and $F_2$ will be about a second and a few milliseconds respectively (by extrapolating the results in \cite{fairMPCCCS17}). MPC overhead will increase the overall operational latency of $\mathsf{MPHTLC}$, but not by a significant amount. With increase in number of participants, we expect that the runtime operation of $F_1$ and $F_2$ will increase, but only by a fraction of one second as suggested by Choudhuri et al. \cite{fairMPCCCS17}.

\begin{table}[h]
\caption{Latency for $\mathsf{MPHTLC}$ operations}
\label{tab:mphtlc-results-latency}
\begin{tabular}{|l|c|c|}
\hline
\textbf{Operation}  & \textbf{Fabric 2P (s)} & \textbf{Corda 2P(s)} \\ \hline \hline
Locks   &   2.124   &   7.486   \\ \hline
Queries &   0.019   &   0.366   \\ \hline
Claims  &   2.121   &   5.898   \\ \hline
Unlocks &   2.128   &   5.892   \\ \hline               
\end{tabular}
\end{table}


\section{Related Work}
\label{sec:relatedWork}
Atomic cross-chain swaps between a pair of mutually distrusting parties without using trusted intermediaries a well-studied problem~\cite{NolanSwap13}. The model and solution have been extended to multiple parties in multiple distinct ledgers~\cite{HPODC18, HSLVLDB19}. Efforts have been made to add robustness to crash faults and address vulnerabilities in HTLC-like solutions~\cite{ZAA19, ZAAVLDB20, XHPODC21}. But all of these works assume singly-owned assets and a single pair of assets across a bilateral link; none of these works consider co-ownerships or multiple assets on a ledger like we do. Therefore, none of the proposed solutions can directly work for our setting, nor can the graph model in \cite{HPODC18} be directly extended to model our scenarios. However, the three attributes of a cross-chain atomic swap protocol defined by Herlihy in \cite{HPODC18} in terms of the guarantees that one such protocol offers are also applicable to our proposed $\mathsf{MPHTLC}$ protocol. 
There also exist several $\mathsf{HTLC}$ implementations (\cite{BIPS17, Decred17, CSGIT21}), but none support scenarios involving co-owned assets.

Some of these works address threats that are orthogonal to the those we address. E.g., in \cite{XHPODC21}, Xue and Herlihy handle {\it sore loser} (or \textit{lockup griefing} attacks~\cite{Arwen2020}) in {\it n-party swaps}, where one party decides to halt participation midway, leaving the other party's assets locked up for a long duration. Our contribution, distinct from theirs, is to extrapolate the basic two-party HTLC model to handle multiple assets and co-owners and provide a solution that ensures atomicity and asset integrity. $\mathsf{MPHTLC}$ is susceptible to sore loser attacks too, but Xue and Herlihy's technique for singly-owned assets (associating premiums to escrows and paying these premiums to counterparties upon refunding)~\cite{XHPODC21} will be equally effective for $\mathsf{MPHTLC}$. Analysis and implementation of $\mathsf{MPHTLC}$ handling sore loser attacks is beyond the scope of this paper though. As an aside, this also allows us to sidestep the question of external market prices of assets impacting the exchange protocol.

In literature, we find other game-theoretic analyses of $\mathsf{HTLC}$-based cross-chain atomic swaps~\cite{BMPSICDCS20, XADICDCS2021} as well as DLT-agnostic and cryptocurrency-agnostic techniques for universal atomic swaps using transaction signature verification~\cite{TMMSP2022}. But none of these works analyze atomic swaps involving assets co-owned by a group of mutually distrusting parties, as we do. 
Uniquely, our work prevents co-owners of an asset swapping it for another asset without the consent of all the co-owners of that asset.


\section{Conclusion and Future Work}
\label{sec:conclusion}
Cross-chain atomic swaps and the $\mathsf{HTLC}$ protocol have been studied in recent years. In this paper, we have presented a general asset exchange model ($\mathcal{GAE}$) whereby assets can be co-owned and multiple assets can be simultaneously exchanged. We have shown that $\mathsf{HTLC}$ cannot be applied directly to fulfil $\mathcal{GAE}$ exchanges because of potential for collusion and fraud. We have presented a solution, $\mathsf{MPHTLC}$ (Multi-Party Hash Time Locked Contract), for $\mathcal{GAE}$, and analyzed its correctness and atomicity properties. We have demonstrated how the protocol can be easily implemented in the Weaver interoperability framework. We intend to demonstrate the practical worth of $\mathsf{MPHTLC}$ in DeFi scenarios by implementing our solution contracts on a permissionless network like Ethereum. Lastly, our conjecture is that $\mathsf{MPHTLC}$ can be extrapolated to multiple ledgers (see Appendix \ref{gas-multiple-ledgers}), and we will investigate this in future work.

\bibliographystyle{ACM-Reference-Format}
\bibliography{aft}

\appendix
\section{Testing asset exchange with co-owners}
\label{sec:appendix1}

We demonstrate asset exchange of a bond in a Fabric network {\em network1} with tokens on Corda network {\em Corda\_Network}, where both the bond and tokens are shared/co-owned.
(the github repository fork containing the source code changes to the {\em Weaver DLT interoperability framework} will be includeded in the camera ready version if this paper gets accepted)

Alice and Bob co-own a bond on {\em network1} while PartyA and PartyB co-own tokens in {\em Corda\_Network}. Here Alice and Bob in Fabric {\em network1} correspond to PartyA ({\em CORDA\_PORT=10006}) and PartyB ({\em CORDA\_PORT=10009}) in {\em Corda\_Network} respectively. Alice's ownership of bond on Fabric network is transferred to Bob (i.e., Bob will be the sole owner of the bond) in exchange for a transfer of Bob's ownership of tokens on Corda network to Alice (i.e., Alice will be the sole owner of the tokens).

Basic capabilities implemented in our modification of Weaver are as follows for an $\mathsf{MPHTLC}$ instance (see the README in the project repository for specific commands to run):
\begin{itemize}[leftmargin=*]
\item Creation of a shared bond on Fabric network
\item Creation of shared tokens on Corda network
\item Exercising $\mathsf{MPHTLC}$ for asset exchange on
shared assets
\end{itemize}

\section{Evaluation of APIs: MPHTLC vs. HTLC}
\label{sec:appendix1a}


We evaluate the overhead for $\mathsf{MPHTLC}$ over conventional $\mathsf{HTLC}$, by considering the transaction latencies for lock/query/claim/unlock of an asset using HTLC APIs and using MPHTLC APIs with single co-owner. We ran our experiments on a Virtual Machine with 8-cores of Intel(R) Xeon(R) Gold 6140 CPU @ 2.30GHz, 16GB RAM, and 64-bit Ubuntu 18.04 as the OS.

Table \ref{tab:mphtlc-vs-htlc-results-latency} reports the latency for locking an asset, querying a locked asset state, claiming a locked asset, and unlocking a locked asset. Latency is computed as the time taken by the respective API (SDK) calls to complete execution. 2nd column (Corda 1P (MPHTLC)) in Table \ref{tab:mphtlc-vs-htlc-results-latency} reports latencies for asset exchange operations with one party co-owning a shared asset before asset exchange and another one party co-owning the same shared asset after asset exchange. And 3rd column (Corda (HTLC)) in Table \ref{tab:mphtlc-vs-htlc-results-latency} reports latencies for asset exchange operations with a party owning a simple asset before asset exchange and a different party owning the same simple asset after asset exchange.
We observe that the times required for the API execution is almost the same for both MPHTLC APIs and HTLC APIs (the asset being exchanged is owned by single owner).

\begin{table}[h]
\caption{Transaction latencies: $\mathsf{HTLC}$ vs. $\mathsf{MPHTLC}$}
\label{tab:mphtlc-vs-htlc-results-latency}
\begin{tabular}{|c|c|c|}
\hline
\textbf{Operation}  & \textbf{Cora 1P (MPHTLC)} & \textbf{Corda (HTLC)} \\ \hline \hline
Locks   &   3.595   &   3.611   \\ \hline
Queries &   0.381   &   0.358   \\ \hline
Claims  &   3.783   &   4.056   \\ \hline
Unlocks &   2.010   &   1.871   \\ \hline               
\end{tabular}
\end{table}

Similar comparison of API execution overhead for MPHTLC vs. HTLC is carried out on Fabric platform as well, and the transaction latencies are in the same range when the assets being exchanged are owned by single owners.

\section{API for Asset Exchange with Co-Owners}
\label{sec:appendix2}

We extended Weaver DLT Interoperability framework to add the functionality of asset exchange with co-owners. Here we present the APIs (i.e. how Layer-2 Application will interact with Layer-1) for lock, claim and reclaim (this doesn't include other changes in interoperation chain code for Fabric and interoperation CorDapp for Corda that we have in our $\mathsf{MPHTLC}$ implementation). The Fabric API is implemented in NodeJS and Corda API is implemented in Kotlin/Java. We also compare these APIs for $\mathsf{MPHTLC}$ with current Weaver's APIs for $\mathsf{HTLC}$.

\subsection{Creation of $\mathsf{MPHTLC}$ Lock}

\begin{itemize}[leftmargin=*]
    \item Fabric: 
    \begin{verbatim}
createSharedHTLC = async(
    contract: Contract, assetType: string,
    assetID: string, lockerECert: string,
    recipientECert: string, hashPreimage: string,
    hashValue: string, expiryTimeSecs: number,
    timeoutCallback: (c: Contract, t: string, 
        i: string, l: string, r: string, 
        p: string, v: string) => any)   \end{verbatim}
    Here $lockerECert$ and $recipientECert$ are comma separated PEM certs of lockers and recipients respectively. That's the only difference as compared to existing $createHTLC$ API in Weaver for Fabric.
    \item Corda: 
    \begin{verbatim} 
createHTLC(
    proxy: CordaRPCOps, assetType: String,
    assetId: String, recipientParty: String,
    hashBase64: String, expiryTimeSecs: Long,
    timeSpec: Int, getAssetStateAndRefFlow: String,
    deleteAssetStateCommand: CommandData, 
    issuer: Party, observers: List<Party>,
    observers: List<Party>, coOwners: List<Party>)      \end{verbatim}
        Here the difference from existing $\mathsf{HTLC}$ API is that $coOwners$ are passed as a list of Corda RPC Parties, in contrast to caller being assumed as sole owner/locker in Weaver $\mathsf{HTLC}$ for Corda.
\end{itemize}

\subsection{$\mathsf{MPHTLC}$ Claim of Locked Asset}

\begin{itemize}[leftmargin=*]
    \item Fabric: 
    \begin{verbatim}
claimSharedAssetInHTLC = async (
    contract: Contract, assetType: string,
    assetID: string, lockerECert: string,
    recipientECert: string, hashPreimage: string)
\end{verbatim} 
    Here $lockerECert$ and $recipientECert$ are comma separated PEM certs of lockers and recipients respectively. That's the only difference as compared to existing $claimAssetInHTLC$ API in Weaver for Fabric.
    \item Corda: There's no change in the API.
\end{itemize}

\subsection{$\mathsf{MPHTLC}$ Unlock of Locked Asset}

\begin{itemize}[leftmargin=*]
    \item Fabric:
    \begin{verbatim}
reclaimSharedAssetInHTLC = async (
    contract: Contract, assetType: string,
    assetID: string, lockerECert: string,
    recipientECert: string)
\end{verbatim}
    Here $lockerECert$ and $recipientECert$ are comma separated PEM certs of lockers and recipients respectively. That's the only difference as compared to existing $reclaimAssetInHTLC$ API in Weaver for Fabric.
    \item Corda: There's no change in the API.
\end{itemize}

\subsection{Query on $\mathsf{MPHTLC}$ Locked State}

\begin{itemize}[leftmargin=*]
    \item Fabric:
    \begin{verbatim}
isSharedAssetLockedInHTLC = async (
    contract: Contract, assetType: string,
    assetID: string, recipientECert: string,
    lockerECert: string)
\end{verbatim} 
    Here $lockerECert$ and $recipientECert$ are comma separated PEM certs of lockers and recipients respectively. That's the only difference as compared to existing $isAssetLockedInHTLC$ API in Weaver for Fabric.
    \item Corda: There's no change in the API.
\end{itemize}

Note that, collecting the consent/signature from the co-owners is achieved in Corda via the transfer of flow session to all the required signers in an automated fashion during the execution of any of these transactions. On the other hand, in Fabric, we don't have this facility to collect the signatures from the co-owners automatically. However, one way to implement this is by collecting the signatures on the hash value (of the secret used for locking) from the required parties manually and validating all the required signatures inside the interoperation chaincode. In our Fabric implementation, we only validate if the transaction submitter is captured as one of the recipients in the lock state during claim/unlock (and one of the co-owners during lock creation) but do not collect the signatures from all the recipients (due to time constraints) and hence we can observe that the execution times of different transactions are less when compared to that of Corda.


\section{$\mathsf{HTLC}$ augmentations and sample attacks}
\label{appendix:HTLC-augmentations-and-attacks}
\begin{figure*}[ht]
    \centering
    \includegraphics[width=0.8\textwidth]{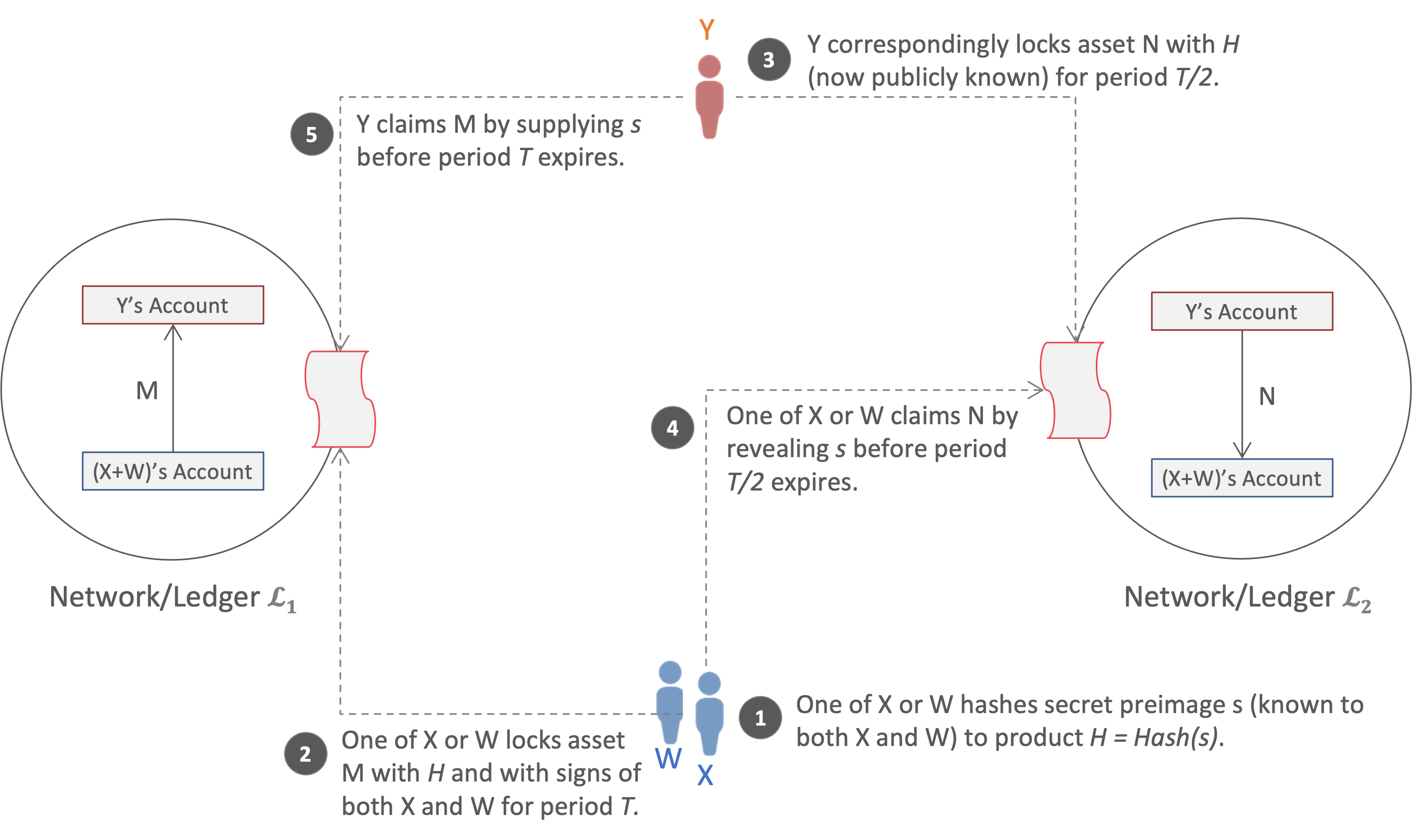}
    \caption{$\mathsf{HTLC}$ Protocol with Multiple Signatures}
    \label{fig:htlc-multi-signs}
\end{figure*}

\begin{figure*}[ht]
    \centering
    \includegraphics[width=0.8\textwidth]{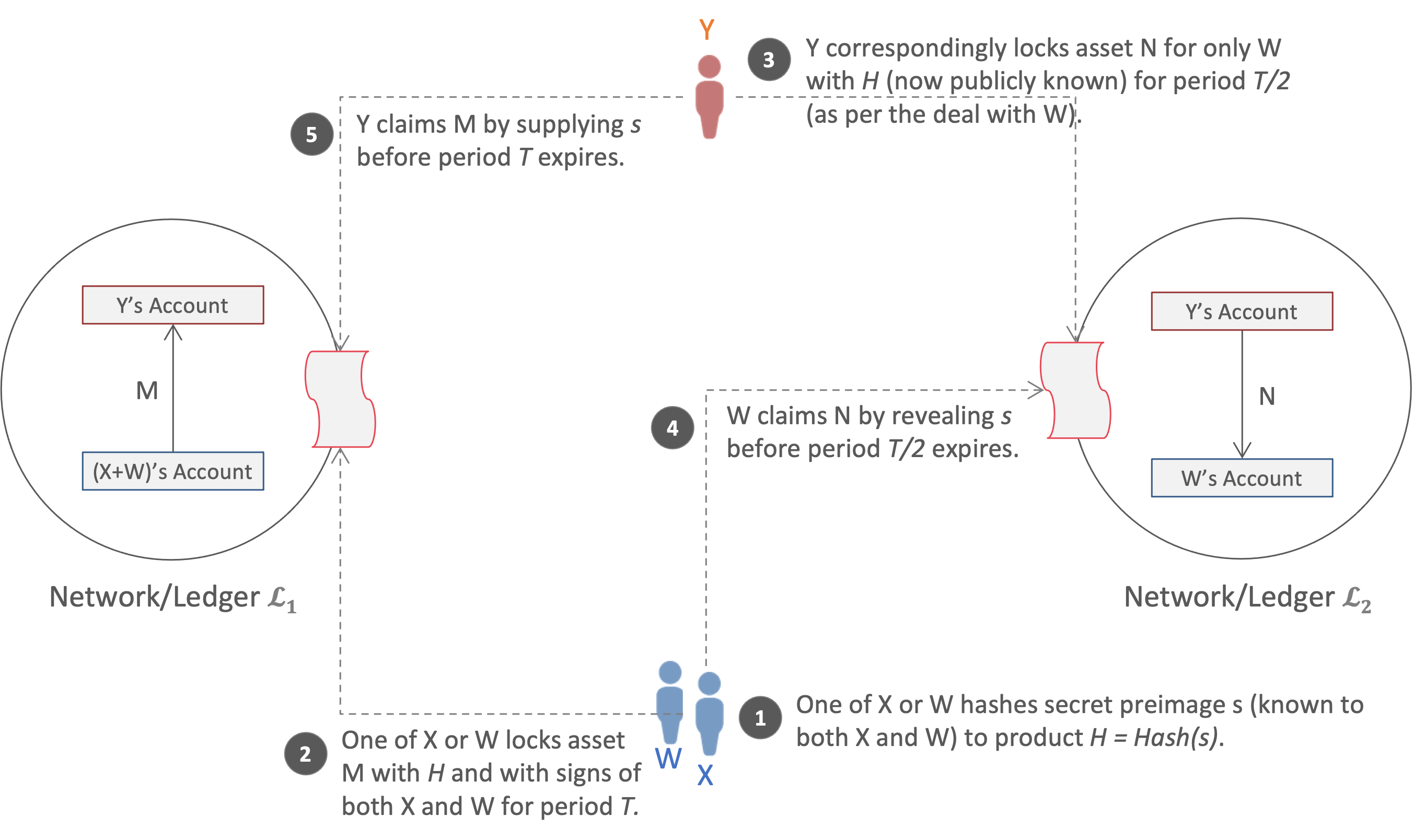}
    \caption{Attack on $\mathsf{HTLC}$ Protocol with Multiple Signatures}
    \label{fig:attack-on-htlc-multi-signs}
\end{figure*}

\begin{figure*}
    \centering
    \includegraphics[width=0.8\textwidth]{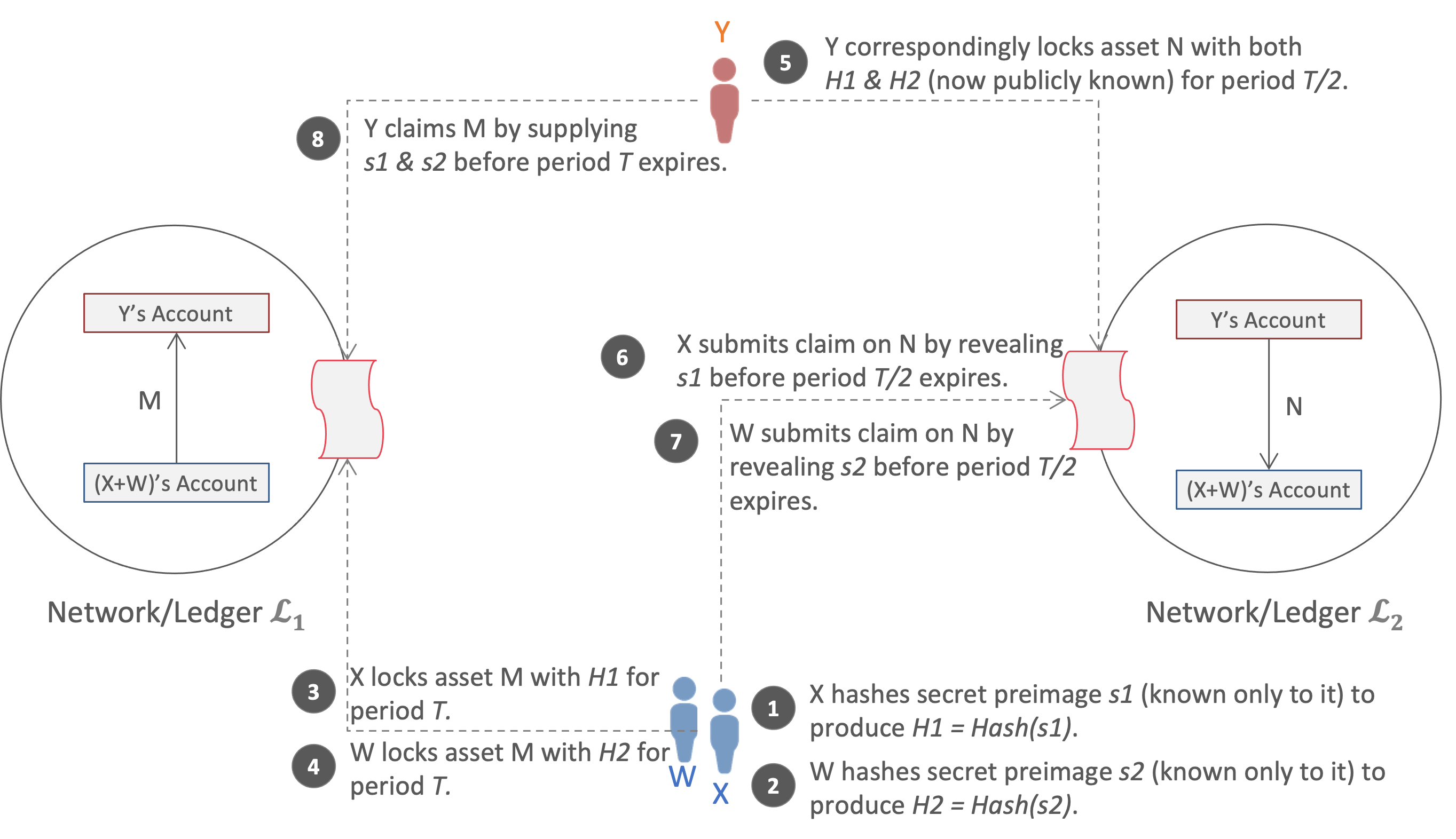}
    \caption{$\mathsf{HTLC}$ Protocol with Multiple Secrets}
    \label{fig:htlc-multi-secrets}
\end{figure*}

\begin{figure*}
    \centering
    \includegraphics[width=0.8\textwidth]{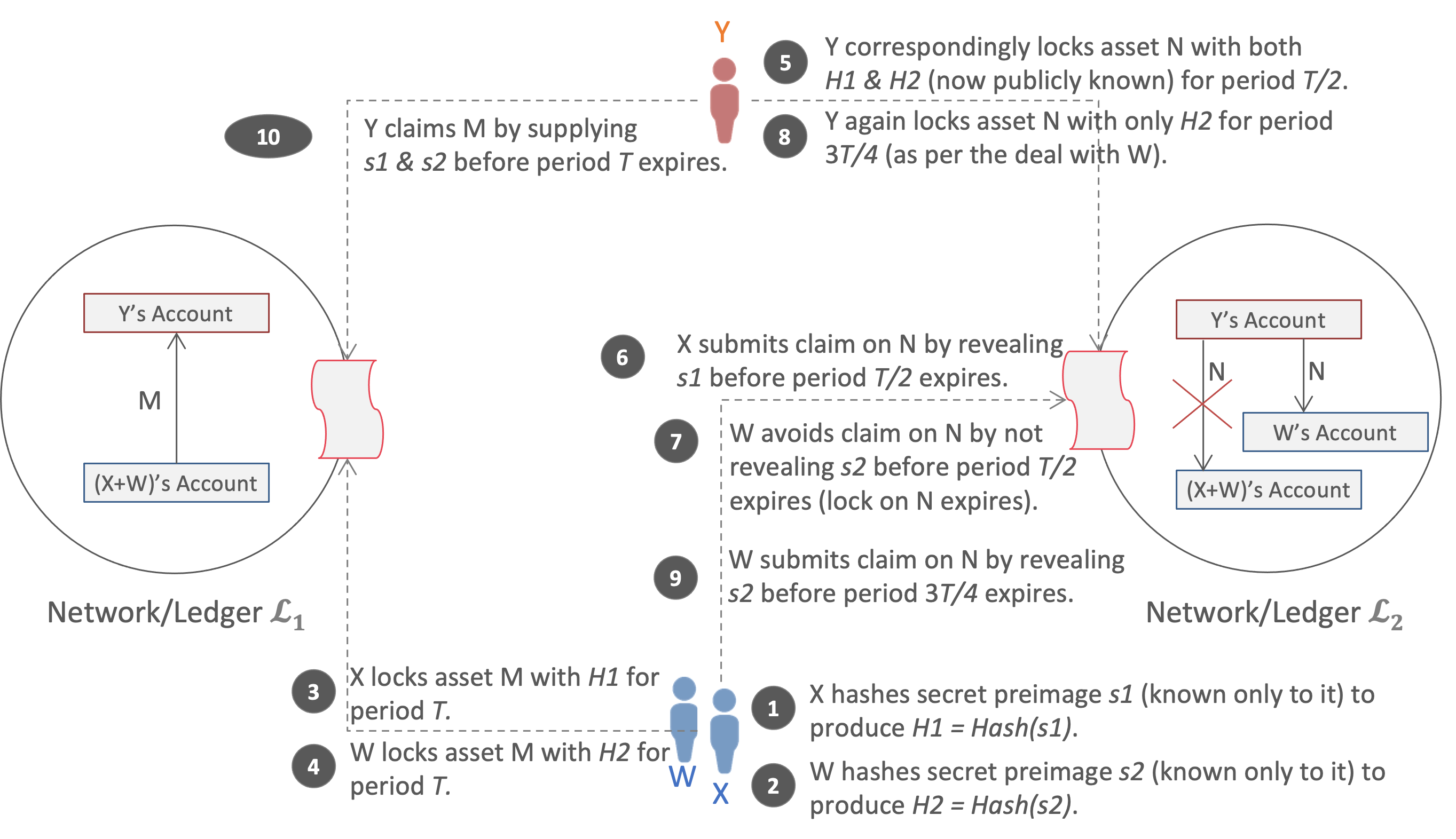}
    \caption{Attack on $\mathsf{HTLC}$ Protocol with Multiple Secrets}
    \label{fig:attack-on-htlc-multi-secrets}
\end{figure*}

Figures \ref{fig:htlc-multi-signs} and \ref{fig:htlc-multi-secrets} represent $\mathsf{HTLC}$ augmentations for shared assets, and figures \ref{fig:attack-on-htlc-multi-signs} and \ref{fig:attack-on-htlc-multi-secrets} represent sample attacks on these modified $\mathsf{HTLC}$ workflows respectively.

\section{$\mathsf{MPHTLC}$ workflow instance}
\label{appendix:MPHTLC-workflow-instance}
\begin{figure*}[ht]
    \centering
    \includegraphics[width=1.4\columnwidth]{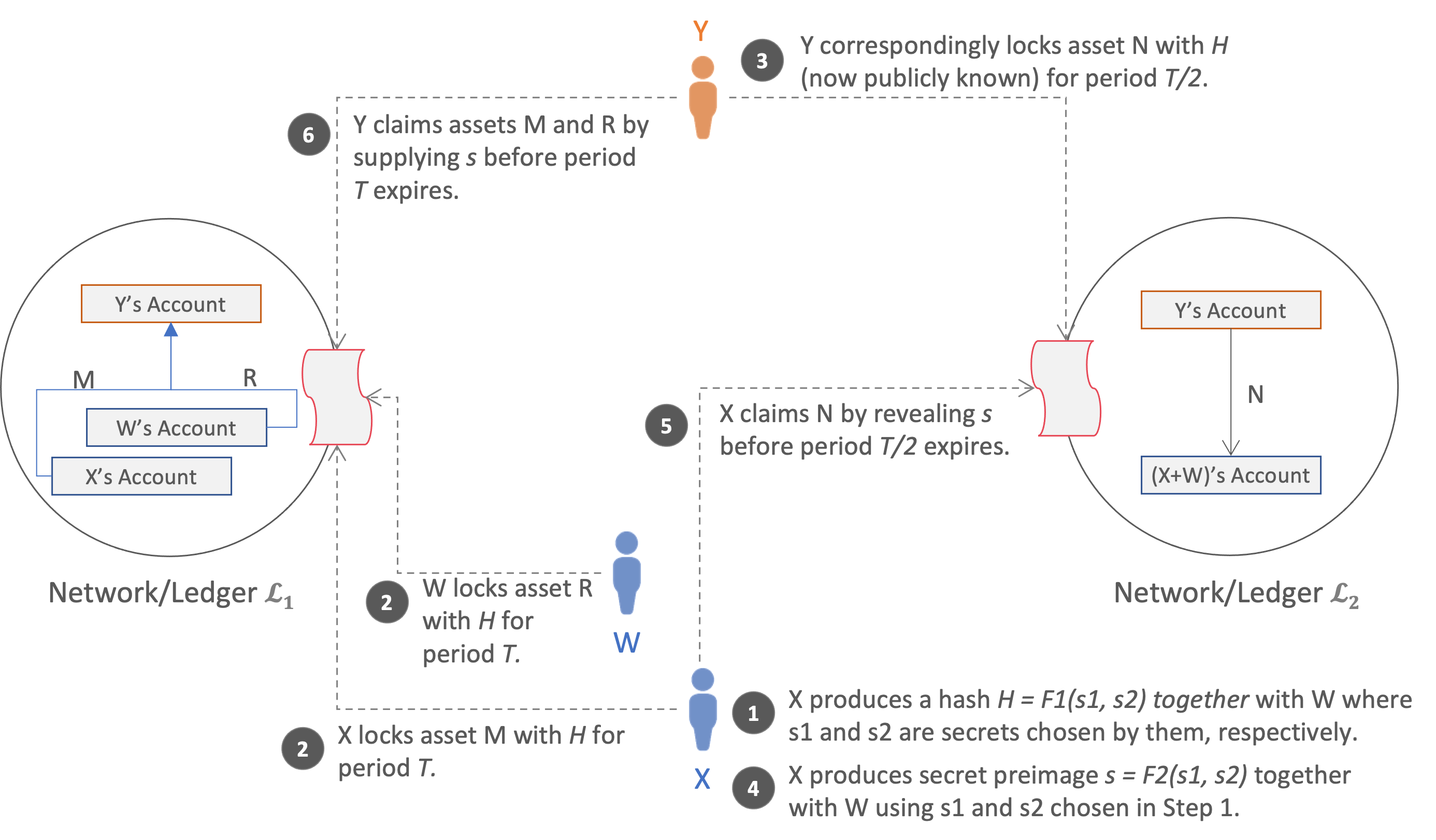}
    \caption{Multi-Party Hash Time Locked Contract ($\mathsf{MPHTLC}$) Protocol}
    \label{fig:mphtlc-maas}
\end{figure*}

Sample $\mathsf{MPHTLC}$ workflow is captured in figure \ref{fig:mphtlc-maas}.

\section{$\mathsf{GAE}$ model with more than two ledgers}
\label{gas-multiple-ledgers}
\begin{figure*}[ht]
    \centering
    \includegraphics[width=1.4\columnwidth]{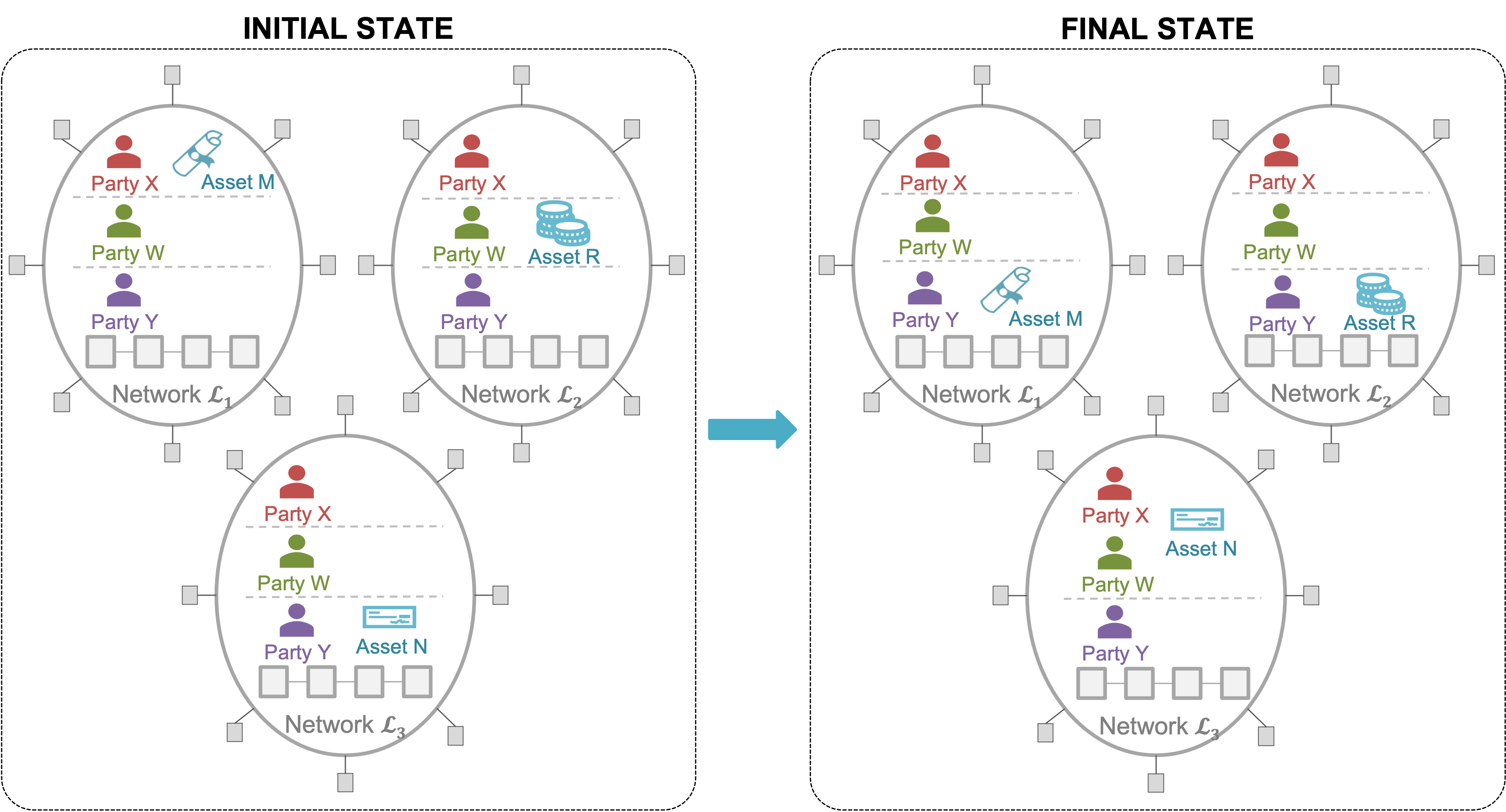}
    \caption{Cross-chain atomic swaps with more than two ledgers}
    \label{fig:assetexchange-variation-nledgers}
\end{figure*}

The discussion till now has focused on parties present in two ledgers. The $\mathcal{GAE}$ model naturally extends to the multi-ledger scenario where more than two ledgers are involved. Consider the asset exchange scenario in Figure \ref{fig:assetexchange-variation-nledgers}. This is similar to the scenario introduced in Figure \ref{fig:asset-exchange-variation-2}. The only difference here is that the assets $\mathcal{W}$ and $\mathcal{R}$ reside in different ledgers. 

Our $\mathsf{MPHTLC}$ protocol can be directly extended to support this multi-ledger scenario. The change from the $\mathsf{MPHTLC}$ protocol in Figure~\ref{fig:mphtlc-maas} is to have the locks and claims on the assets applied in the respective ledgers.
We defer to future work to formally write down the multi-ledger $\mathcal{GAE}$ model, thoroughly investigate the base scenarios and prove that the base cases and the solution to them are complete for a multi-ledger $\mathcal{GAE}$.


\section{Corda Network Architecture in Weaver}
\label{weaver-arch-corda}

\begin{figure*}[ht]
    \centering
    \includegraphics[width=0.99\columnwidth]{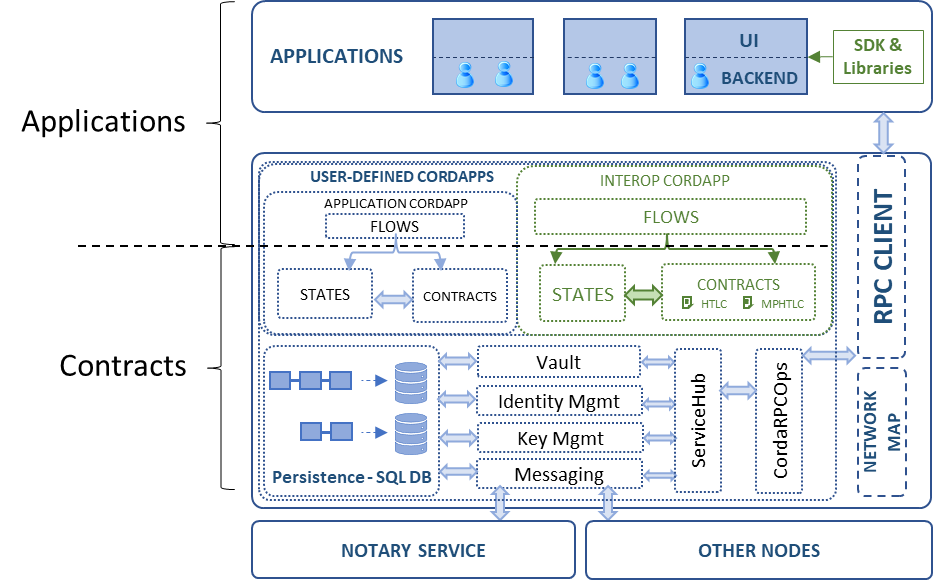}
    \caption{Corda Network and Decentralized App Architecture}
    \label{fig:implementation-corda}
\end{figure*}

Figure \ref{fig:implementation-corda} describes the architecture of the decentralized apps in a Corda network.

\end{document}